\documentclass[journal]{IEEEtran}

\ifCLASSINFOpdf
\else
   \usepackage[dvips]{graphicx}
\fi
\usepackage{url}

\usepackage{etoolbox}
\robustify\bfseries
\usepackage{graphicx}
\usepackage{amsmath,amssymb,graphicx}
\usepackage[]{algorithm2e}
\usepackage{tabularx}
\usepackage{wasysym}
\usepackage{color,soul}
\usepackage{cite}
\usepackage{balance}
\usepackage{breqn}
\usepackage{makecell}
\usepackage{booktabs}
\usepackage{multirow}
\usepackage{siunitx}
\usepackage{caption}
\usepackage{xcolor}
\usepackage{hyperref}
\def\H{{\mathsf H}}
\def\T{{\mathsf T}}
\def\CC{{\mathbb C}}
\def\RR{{\mathbb R}}
\usepackage{caption}
\usepackage{arydshln}
\usepackage{enumitem}

\makeatletter
\newcommand*{\bigs}[1]{{\hbox{$\left#1\vbox to9\p@{}\right.\n@space$}}}
\makeatother

\usepackage{amssymb}%
\usepackage{pifont}%
\newcommand{\cmark}{\ding{51}}%
\newcommand{\xmark}{\ding{55}}%

\newcommand{\ZQHL}[1]{{#1}}

\begin{document}

\title{TF-GridNet: Integrating Full- and Sub-Band Modeling for Speech Separation}

\author{Zhong-Qiu Wang, Samuele Cornell, Shukjae Choi, Younglo Lee, Byeong-Yeol Kim, and Shinji Watanabe
\thanks{Manuscript received on Nov. 22, 2022; revised on Jul. 9, 2023.}
\thanks{
Z.-Q. Wang and S. Watanabe are with the Language Technologies Institute, Carnegie Mellon University, Pittsburgh, PA 15213, USA (e-mail: wang.zhongqiu41@gmail.com; shinjiw@cmu.edu).}
\thanks{
S. Cornell is with Department of Information Engineering, Università Politecnica delle Marche, Acona, Italy (e-mail: cornellsamuele@gmail.com).}
\thanks{
S. Choi, Y. Lee and B.-Y. Kim are with Hyundai Motor Group and 42dot Inc., Seoul, Korea (e-mail: sjchoi.hmg@gmail.com; 
\{younglo.lee, byeongyeol.kim\}@42dot.ai).}
}

\markboth{}
{Shell \MakeLowercase{\textit{et al.}}: Bare Demo of IEEEtran.cls for IEEE Journals}
\maketitle

\begin{abstract}

We propose TF-GridNet for speech separation.
The model is a novel
deep neural network (DNN) integrating full- and sub-band modeling in the time-frequency (T-F) domain.
It stacks several
blocks, each consisting of an intra-frame full-band module, a sub-band temporal module, and a cross-frame self-attention module.
It is trained to perform complex spectral mapping, where the real and imaginary (RI) components of input signals are stacked as features to predict target RI components.
We first evaluate it on monaural anechoic speaker separation.
Without using data augmentation and dynamic mixing, it obtains a state-of-the-art 23.5 dB improvement in scale-invariant signal-to-distortion ratio (SI-SDR) on WSJ0-2mix, a standard dataset for two-speaker separation.
To show its robustness to noise and reverberation, we evaluate it on monaural reverberant speaker separation using the SMS-WSJ dataset and on noisy-reverberant speaker separation using WHAMR!, and obtain state-of-the-art performance on both datasets.
We then extend TF-GridNet to multi-microphone conditions through multi-microphone complex spectral mapping, and integrate it into a two-DNN system with a beamformer in between (named as MISO-BF-MISO in earlier studies), where the beamformer proposed in this paper is a novel multi-frame Wiener filter computed based on the outputs of the first DNN.
State-of-the-art performance is obtained on the multi-channel tasks of SMS-WSJ and WHAMR!.
Besides speaker separation, we apply the proposed algorithms to speech dereverberation and noisy-reverberant speech enhancement.
State-of-the-art performance is obtained on a dereverberation dataset and on the dataset of the recent L3DAS22 multi-channel speech enhancement challenge. 

\end{abstract}

\begin{IEEEkeywords}
Complex spectral mapping, acoustic beamforming, full- and sub-band integration, speech separation.
\end{IEEEkeywords}

\IEEEpeerreviewmaketitle

\vspace{-0.1cm}
\section{Introduction}

\IEEEPARstart{D}{eep} learning has dramatically advanced talker-independent speaker separation in the past decade~\cite{WDLreview}, especially since deep clustering~\cite{Hershey2016} and permutation invariant training (PIT)~\cite{Kolbak2017} successfully addressed the label permutation problem.
Early studies train DNNs for magnitude estimation, with or without estimating phase~\cite{Isik2016, Wang2018AlternativeObejectives, WZQe2eMISI2018, Wang2019Trigonometric}.
Subsequent studies carry out separation in the complex T-F domain via complex ratio masking~\cite{Liu2019DeepCASA} or in the time domain via TasNets~\cite{Luo2017TasNet, Luo2018TasNetRealTime, Luo2019}.
Since 2019, Conv-TasNet and its variants~\cite{Luo2019, Lam2020MBT, Shi2019FurcaNeXt, Tzinis2020, Luo2020, Nachmani2020, Chen2020DPTnet, Zhu2021, Subakan2021, Lam2021, Zeghidour2020, Qian2022, Rixen2022, Rixen2022QDPN}, featuring advanced DNN architectures with learned encoder-decoder modules operating on very short windows of signals for end-to-end masking based separation, have gradually become the most popular and dominant approach for speaker separation in anechoic conditions.
Their performance on the standard WSJ0-2mix benchmark~\cite{Hershey2016} has reached an impressive SI-SDR improvement (SI-SDRi) of 22.1 dB~\cite{Rixen2022QDPN}.

In the meantime, T-F domain models, which usually use larger window and hop sizes, have been largely under-explored and under-represented in speaker separation in anechoic conditions.
\ZQHL{Recently, TFPSNet \cite{Yang2022TFPSNet} reported a strong SI-SDRi of 21.1 dB on WSJ0-2mix, which is comparable to the top results achievable by modern time-domain models.}
It leverages a modern dual-path architecture, following DPRNN \cite{Luo2020} and DPTNet \cite{Chen2020DPTnet}, but applies the architecture on complex T-F spectrogram \cite{Le2021DPCRN, Dang2022DPTFSNet} by using the transformer module proposed in DPTNet \cite{Chen2020DPTnet} to model spectro-temporal information.
\ZQHL{Although TFPSNet operates in the T-F domain \cite{Yang2022TFPSNet}}, it closely follows the encoder-separator-decoder scheme \cite{Luo2019} widely-used in TasNets and its performance, even with a modern DNN architecture, is still much lower than contemporary time-domain models \cite{Qian2022, Rixen2022, Rixen2022QDPN}.

In this context, for anechoic speaker separation our preliminary version~\cite{Wang2022GridNet} of this paper made the following contributions to improve complex T-F domain approaches:
\begin{itemize}[leftmargin=*,noitemsep,topsep=0pt]
\item We \ZQHL{proposed} to use complex spectral mapping for speaker separation in anechoic conditions.
Complex spectral mapping~\cite{Williamson2016, Fu2017, Tan2020, Wang2020CSMDereverbJournal, Wang2020chime, Wang2020dMCCSMconference, Wang2020css, Wang2021LowDistortion, Wang2021FCPjournal, Tan2022NSF, Wang2021FCPwaspaa}, which predicts target RI components based on the RI components of input signals, has shown strong potential on noisy-reverberant speech separation when combined with modern DNN architectures and loss functions, exhibiting strong robustness to noise and reverberation in both single- and multi-microphone conditions.
Its potential on anechoic speaker separation, however, has not been studied, especially in an era when time-domain models, which perform masking in a learned filterbank domain, have become so popular and dominant on this task.
This paper is the first study to explore this direction for monaural, anechoic speaker separation.
\item We \ZQHL{proposed} a novel DNN architecture named TF-GridNet for speech separation.
It operates in the complex T-F domain to model speech spectrograms in a grid-like manner.
Based on an improved TFPSNet~\cite{Yang2022TFPSNet}, we add a cross-frame self-attention path for dual-path models to leverage global information across frames;%
\item Building upon the SI-SDR loss~\cite{Luo2019, LeRoux2019}, we \ZQHL{proposed} to add a novel loss term to encourage estimated sources to add up to the mixture. We also combine this loss term with loss functions other than SI-SDR.
\end{itemize}
Without using any data augmentation and dynamic mixing, on WSJ0-2mix~\cite{Hershey2016} our best model obtains 23.5 dB SI-SDRi, clearly better than the previous best (at 22.1 dB)~\cite{Rixen2022QDPN}.

However, our preliminary study~\cite{Wang2022GridNet} does not show the potential of TF-GridNet for speech separation in noisy-reverberant conditions and it lacks an extension to multi-channel conditions.
To address the first problem, we evaluate TF-GridNet on monaural reverberant speaker separation using the SMS-WSJ dataset~\cite{Drude2019} and monaural noisy-reverberant speaker separation using WHAMR!~\cite{Maciejewski2020}.
To address the second problem, we integrate TF-GridNet with a MISO-BF-MISO approach~\cite{Wang2020dMCCSMconference, Wang2020css, Wang2021LowDistortion}, which sandwiches a beamformer with two multi-channel-input single-channel-output (MISO) DNNs, with the beamformer computed based on the output of the first DNN and the second DNN performing post-filtering.
In our recent work~\cite{Lu2022}, we follow this MISO-BF-MISO approach and stack two TCN-DenseUNets with a novel multi-channel multi-frame Wiener filter (MFWF) in between.
The TCN-DenseUNet~\cite{Wang2020dMCCSMconference, Wang2020CSMDereverbJournal, Wang2020chime, Wang2020css, Wang2021LowDistortion, Wang2021FCPjournal, Tan2022NSF, Wang2021FCPwaspaa} is a strong, representative model adopted in many previous complex spectral mapping studies, and the MFWF is computed based on both DNN-estimated target magnitude and phase, and leverages both future and past frames for sub-band linear filtering.
This solution won the recent L3DAS22 3D speech enhancement challenge~\cite{Guizzo2022L3DAS}, which attracted 17 submissions.
In this paper, a major difference from~\cite{Lu2022} is that we replace the TCN-DenseUNet with the newly-proposed TF-GridNet by modifying TF-GridNet for multi-microphone complex spectral mapping~\cite{Wang2020dMCCSMconference, Wang2020css, Wang2021LowDistortion}, and we observe large improvement over~\cite{Lu2022} and many other strong multi-channel systems.
Both TF-GridNet and MISO-BF-MISO can be understood from the perspective of integrated full- and sub-band modeling, either inside TF-GridNet or outside through beamforming and post-filtering.

State-of-the-art performance is achieved on four major speech separation tasks, including reverberant speaker separation, noisy-reverberant speaker separation, speech dereverberation and noisy-reverberant speech enhancement, showing the effectiveness of the proposed algorithms at single- and multi-channel separation.
In our experiments, for each task we strive to use public datasets with strong results published by previous studies.
A sound demo is available online.\footnote{See \url{https://zqwang7.github.io/demos/TF-GridNet-demo/index.html}.}
\ZQHL{We have released the code of TF-GridNet in the ESPnet-SE++ toolkit~\cite{Lu2022ESPNetSE++}.\footnote{See 
\url{https://github.com/espnet/espnet/pull/5395}}}

\section{System Overview}\label{systemoverview}

\subsection{Physical Model and Objective}

For an $N$-sample, $C$-speaker mixture signal recorded by a $P$-microphones array in a noisy-reverberant setting, at sample $n$ the physical model describing the relationship between the mixture $\mathbf{y}[n] \in \RR^P$, reverberant non-target signals $\mathbf{v}[n]\in \RR^P$, and dry source signal $(o(c))[n]\in \RR$, direct-path signal $(\mathbf{s}(c))[n]\in \RR^P$ and reverberation $(\mathbf{h}(c))[n]\in \RR^P$ of speaker $c$ can be formulated in the time domain as
\begin{align} 
	\mathbf{y}&[n] = \sum\nolimits_{c=1}^C \Big(o(c) * \mathbf{r}(c)\Big)[n] + \mathbf{v}[n] \nonumber \\
&= \sum\nolimits_{c=1}^C \Big(\left(o(c) * \mathbf{r}^d(c)\right)[n] + \left(o(c) * \mathbf{r}^{e+l}(c)\right)[n]\Big) + \mathbf{v}[n] \nonumber \\
&= \sum\nolimits_{c=1}^C \Big(\left(\mathbf{s}(c)\right)[n] + \left(\mathbf{h}(c)\right)[n]\Big) + \mathbf{v}[n], \label{eq:phymodel_time}
\end{align}
where $*$ is the linear convolution operator, and the $P$-channel room impulse response (RIR) of speaker $c$, $\mathbf{r}(c)$, can be decomposed into the RIR of the direct-path signal, $\mathbf{r}^d(c)$, and that of early reflections and late reverberation combined, $\mathbf{r}^{e+l}(c)$.
In the short-time Fourier transform (STFT) domain, the physical model is formulated as
\begin{align} 
	\mathbf{Y}(t,f) &= \sum\nolimits_{c=1}^C \Big(\mathbf{S}(c, t,f)+\mathbf{H}(c,t,f)\Big)+\mathbf{V}(t,f), \label{eq:phymodel_freq}
\end{align}
where $t$ indexes $T$ frames, $f$ indexes $F$ frequencies, and $\mathbf{Y}(t,f)$, $\mathbf{V}(t,f)$, and $\mathbf{S}(c,t,f)$ and $\mathbf{H}(c,t,f)\in \CC^P$ respectively denote the STFT vectors of the mixture, non-target signals, and the direct-path signal and reverberation of speaker $c$.
The corresponding spectrograms are denoted by $\mathbf{Y}$, $\mathbf{V}$, $\mathbf{S}(c)$, and $\mathbf{H}(c)$.
This formulation covers all the tasks we consider:
\begin{itemize}[leftmargin=*,noitemsep,topsep=0pt]
\item For monaural, anechoic speaker separation, $C > 1$, $P=1$ and there is no $\mathbf{H}$ and $\mathbf{V}$;
\item For reverberant speaker separation, $C > 1$ and $\mathbf{V}$ is a weak stationary noise (e.g., microphone sensor noise); 
\item For noisy-reverberant speaker separation, $C > 1$ and $\mathbf{V}$ consists of challenging non-stationary noises;
\item For speech dereverberation, $C=1$ and $\mathbf{V}$ is a weak stationary noise;
\item For noisy-reverberant speech enhancement, $C=1$ and $\mathbf{V}$ contains challenging non-stationary noises.
\end{itemize}
Given a single- or multi-channel mixture, we aim at reconstructing the direct-path signal of each speaker at a reference microphone $q$ (i.e., $s_q(c)$).
This requires us to not only remove noise and reverberation but also separate the speakers if there are more than one.
For all the tasks, we assume that the maximum number of speakers in each mixture is known, and that the array geometry is fixed between training and testing.

\begin{figure}
  \centering  
  \includegraphics[width=7.5cm]{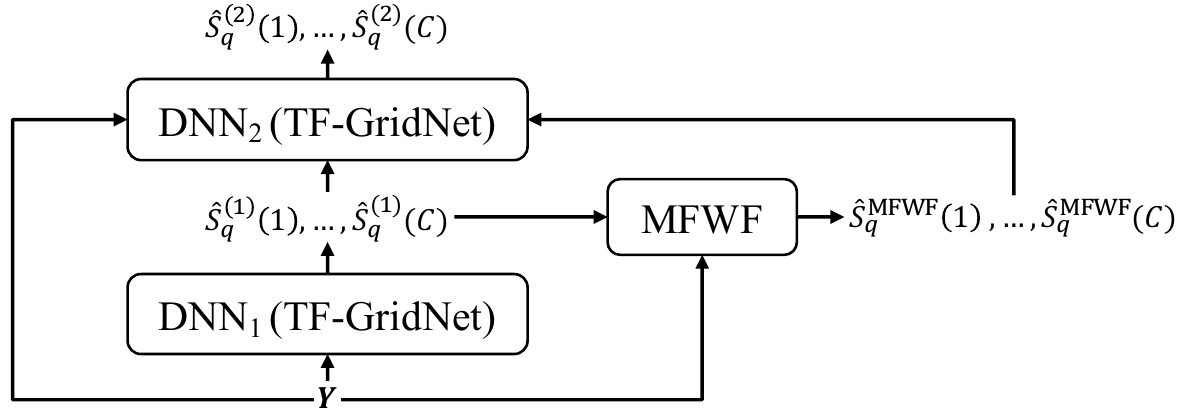}
  \vspace{-0.1cm}
  \caption{System overview.}
  \label{system_overview}
  \vspace{-0.6cm}
\end{figure}

\begin{figure*}[h!]
  \centering  
  \includegraphics[width=18cm]{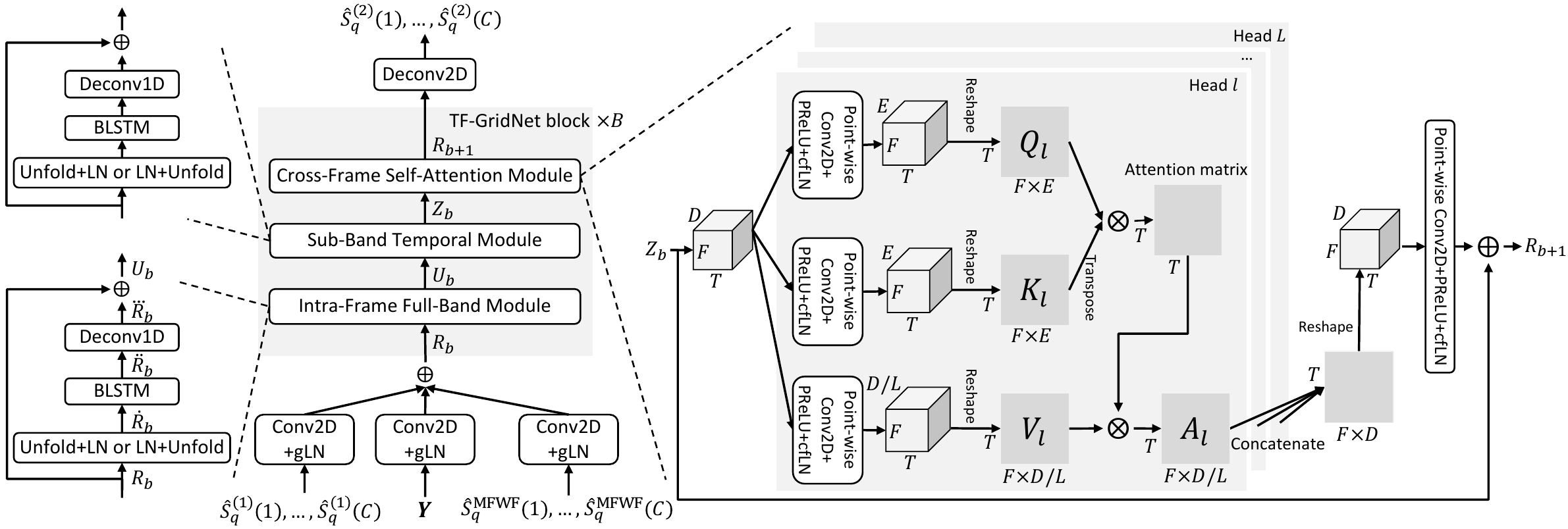}
  \vspace{-0.1cm}
  \caption{Proposed TF-GridNet based DNN$_2$.}
  \label{GridNet_overview}
  \vspace{-0.5cm}
\end{figure*}

\subsection{Approach Outline}

Our system (see Fig.~\ref{system_overview}) operates in the complex T-F domain.
It follows a two-DNN approach named MISO-BF-MISO~\cite{Wang2021LowDistortion, Wang2020css, Wang2021FCPjournal}, where DNN$_1$ first produces an initial estimate for each target source, the initial estimate is then used to compute a sub-band linear filter (in this paper a multi-frame Wiener filter) for each source, and DNN$_2$ takes in the mixture, the outputs of DNN$_1$, and the linear-filtering results for post-filtering.
In our experiments, DNN$_1$ and DNN$_2$ are trained sequentially rather than jointly.
After DNN$_1$ is trained, we use it to generate an initial estimate $\hat{S}_q^{(1)}(c)$ and compute a sub-band linear filtering result $\hat{S}_q^{\text{MFWF}}(c)$ for each speaker $c$, and feed them and $\mathbf{Y}$ to DNN$_2$ to further predict target speech (denoted as $\hat{S}_q^{(2)}(c)$).
The superscripts in $\hat{S}_q^{(1)}(c)$ and $\hat{S}_q^{(2)}(c)$ denote which of the two DNNs produces the estimate.
Following~\cite{Wang2020css, Wang2021FCPjournal, Wang2021FCPwaspaa}, for speaker separation DNN$_1$ is trained with utterance-wise PIT~\cite{Kolbak2017} but DNN$_2$ is trained in an enhancement way (i.e, predicting all the speakers but not using PIT), since the label-permutation problem has been addressed by DNN$_1$.
For monaural, anechoic speaker separation, we only train DNN$_1$, without using linear filtering and DNN$_2$.

\section{TF-GridNet}\label{proposedalgorithm}

Fig.~\ref{GridNet_overview} illustrates the proposed TF-GridNet for DNN$_2$.
DNN$_1$ has the same architecture but uses only $\mathbf{Y}$ as input.
Both DNNs are trained to perform complex spectral mapping~\cite{Williamson2016, Fu2017, Tan2020, Wang2020CSMDereverbJournal, Wang2020chime, Wang2020dMCCSMconference, Wang2021LowDistortion, Wang2020css, Wang2021FCPjournal, Tan2022NSF, Wang2021FCPwaspaa}, where the RI components of input signals are stacked as input features to predict the RI components of each speaker at the reference microphone $q$, i.e., $S_q(c)$.
Our system is non-causal.
We normalize the sample variance of each input 
mixture to $1.0$ and use the same scaling factor to scale each target source before using them for training. %
This amounts to adjusting the volume of each input mixture to a similar level.

In Fig.~\ref{GridNet_overview}, for each of the three real-valued input features (i.e., \ZQHL{the stacked RI components of} the mixture $\mathbf{Y}$ with shape $2P\times T\times F$, DNN$_1$'s outputs $\hat{S}_q^{(1)}(1),\dots,\hat{S}_q^{(1)}(C)$ with shape $2C\times T\times F$, and MFWF's outputs $\hat{S}_q^{\text{MFWF}}(1),\dots,\hat{S}_q^{\text{MFWF}}(C)$ with shape $2C\times T\times F$), we first use a two-dimensional (2D) convolution (Conv2D) with a $3\times 3$ kernel followed by global layer normalization (gLN)~\cite{Luo2019} to compute a $D$-dimensional embedding for each T-F unit, and then summate the T-F embeddings generated for the three input features, obtaining a tensor with shape $D\times T\times F$.
Next, we feed the tensor to $B$ \ZQHL{stacked} TF-GridNet blocks, each consisting of an intra-frame full-band module, a sub-band temporal module, and a cross-frame self-attention module, to leverage spectral, spatial and temporal information to gradually make the T-F embeddings more discriminative for separation.
After that, a 2D deconvolution (Deconv2D) with $2C$ output channels and a $3\times 3$ kernel followed by linear units is used to obtain the predicted RI components for all the $C$ speakers, and inverse STFT (iSTFT) is applied for signal re-synthesis.
The rest of this section describes the three modules in each TF-GridNet block, and the loss functions.
To avoid confusion, in Table~\ref{summary_hyperparam} we list the hyper-parameters we will use to describe TF-GridNet.

\begin{table}[t]
\scriptsize
\centering
\sisetup{table-format=2.2,round-mode=places,round-precision=2,table-number-alignment = center,detect-weight=true,detect-inline-weight=math}
\caption{\textsc{List of Hyper-Parameters of TF-GridNet.}}
\vspace{-0.2cm}
\label{summary_hyperparam}
\setlength{\tabcolsep}{2pt}
\begin{tabular}{
cc
}
\toprule

Symbols & Description  \\

\midrule

$D$ & Embedding dimension for each T-F unit  \\
$B$ & Number of TF-GridNet blocks  \\
\midrule
$I$ & Kernel size for Unfold and Deconv1D \\
$J$ & Stride size for Unfold and Deconv1D \\
$H$ & Number of hidden units of BLSTMs in each direction  \\
\midrule
$L$ & Number of heads in self-attention  \\
$E$ & \begin{tabular}{@{}c@{}}Number of output channels in point-wise Conv2D to obtain key and query\\tensors in self-attention\end{tabular} \\
\bottomrule
\end{tabular}
\vspace{-0.6cm}
\end{table}

\subsection{Intra-Frame Full-Band Module}

For the intra-frame module, we view the input tensor $R_b \in \RR^{D\times T \times F}$ to the $b^{\text{th}}$ block as $T$ separate sequences, each with length $F$, and use a sequence model to capture the full-band spectral and spatial information within each frame.

In detail, we first use the \textit{torch.unfold} function~\cite{Paszke2019} with kernel size $I$ and stride $J$ to stack nearby embeddings at each step along frequency, after zero-padding the frequency dimension to $F'= \lceil \frac{F-I}{J} \rceil \times J + I$, and then apply layer normalization (LN) along the first dimension, i.e.,
\begin{align}\label{intra-frame-unfold-ln}
\dot{R}_b = \text{LN}\Big(\Big[&\text{Unfold}(R_b[:,t,:]),\nonumber \\
&\text{for}\,\,t=1,\dots,T\Big]\Big) \in \RR^{(I\times D)\times T \times (\frac{F'-I}{J}+1)}.
\end{align}
We denote this order of operations as \textbf{Unfold-LN}.
An alternative is to first perform LN on $R_b$ and then zero-pad and stack nearby embeddings, i.e., 
\begin{align}\label{intra-frame-ln-unfold}
\dot{R}_b = \Big[\text{Un}&\text{fold}\Big(\text{LN}(R_b)[:,t,:]\Big),\nonumber \\
&\text{for}\,\,t=1,\dots,T\Big] \in \RR^{(I\times D)\times T \times (\frac{F'-I}{J}+1)}.
\end{align}
We denote this order as \textbf{LN-Unfold}.
We point out that LN-Unfold uses fewer parameters than Unfold-LN, and, since the \textit{torch.unfold} function creates a view of the input tensor without allocating new memory, LN-Unfold consumes less memory when $I/J > 1$.
Note that our preliminary paper~\cite{Wang2022GridNet} uses Unfold-LN, and this paper proposes LN-Unfold, which leads to slightly better separation.

We then use a \ZQHL{single} bi-direcitonal long short-term memory (BLSTM) with $H$ units in each direction to model inter-frequency information within each frame: %
\begin{align}\label{intra-frame-blstm}
\ddot{R}_b = \Big[\text{BLSTM}\big(&\text{LN}(\dot{R}_b)[:,t,:]\big), \nonumber \\
&\text{for}\,\,t=1,\dots,T\Big]
\in \RR^{2H\times T \times (\frac{F'-I}{J}+1)}.
\end{align}
Note that $J$ can be larger than one so that the sequence length and thus the amount of computation can be reduced.

Next, a one-dimensional deconvolution (Deconv1D) layer with kernel size $I$, stride $J$, input channel $2H$ and output channel $D$ (and without subsequent normalization and non-linearity) is applied to the hidden embeddings of the BLSTM: %
\begin{align}\label{intra-frame-deconv}
\dddot{R}_b = \Big[\text{Deconv1D}(&\ddot{R}_b[:,t,:]), \nonumber \\ &\text{for}\,\,t=1,\dots,T\Big] \in \RR^{D\times T \times F'}.
\end{align}
After removing zero paddings, this tensor is added to the input tensor via a residual connection to produce the output tensor:
\begin{align}\label{intra-frame-residual}
U_b = \dddot{R}_b[:,:,:F] + R_b \in \RR^{D\times T \times F}.
\end{align}

\subsection{Sub-Band Temporal Module}\label{subband_description}

In the sub-band temporal module, the procedure is almost the same as that in the intra-frame full-band module.
The only difference is that the input tensor $U_b \in \RR^{D\times T\times F}$ is viewed as $F$ separate sequences, each with length $T$, and a BLSTM is used to model the temporal information within each frequency.
\ZQHL{Note that the parameters of the BLSTM are shared across all the frequencies.}
The output tensor is denoted as $Z_b \in \RR^{D\times T\times F}$.

\subsection{Discussion on Full- and Sub-Band Modeling }\label{fullsubband_discussion}

In multi-channel conditions, performing sub-band modeling is a reasonable strategy to leverage spatial information afforded by multiple microphones.
The idea is that inter-microphone spatial patterns such as the inter-channel phase differences (IPD) do not change along time for sources that do not move within each utterance, while they usually change with frequency due to the linear phase structure of phase differences
and the effects of phase wrapping (see an example plot of IPD vs. frequency in anechoic conditions in Fig. 3 of~\cite{Wang2018combiningspectralspatial}).
This is partly the reason why many conventional beamforming~\cite{Gannot2017}, dereverberation~\cite{Nakatani2010} and spatial clustering~\cite{Haeb-Umbach2020} algorithms are performed separately within each frequency.
In light of this physical phenomenon, we believe that it intuitively makes sense to perform such a DNN-based sub-band modeling, as the inter-channel phase patterns important for supervised learning are stable and salient within each frequency for each source.
In addition, using a shared DNN block to separately model each sub-band is easier than using a DNN block to simultaneously model all the frequencies, as there are fewer variations to model. 
This echoes the idea of weight sharing, a core concept in convolutional neural networks~\cite{Courville2016}. %

Similarly, in multi-microphone conditions the intra-frame full-band module described in the previous subsection could not only model the full-band, spectral patterns such as the harmonic structure along frequency but also model the gradual changes of inter-microphone phase patterns along frequency (see the helix structure of IPD along frequency in Fig. 3(c) of~\cite{Wang2018combiningspectralspatial}).
We emphasize that the pattern of such gradual changes along frequency exists at every frame where the target source (assumed non-moving) is active.
It is therefore reasonable to run the same BLSTM based full-band module at each frame to model such patterns.

Such sub-band modeling approach could better deal with reverberation.
Since reverberation time (T60) and reverberation patterns vary with frequency~\cite{A.P.Habets2018DereverbBook}, it is reasonable to use sub-band modules in TF-GridNet to separately model each frequency.
In a broader perspective, weighted prediction error (WPE)~\cite{Nakatani2010}, the most popular conventional algorithm for dereverberation, is also performed per-frequency by computing a linear, inverse filter at each frequency (preferably with different number of filter taps at different frequencies~\cite{Nakatani2019ConvBeamformer}) to estimate late reverberation.
There are studies~\cite{Zhao2018cLSTMLateReverb} using a non-linear LSTM to mimic the linear, inverse filtering of WPE, but the LSTM is trained to model all the frequencies simultaneously rather than separately.
We believe that using sub-band DNN modules to mimic sub-band inverse filtering is likely better, because reverberation, at each frequency, can be approximated as a linear convolution of a sub-band filter and the anechoic signal, according to the narrow-band approximation property~\cite{Gannot2017, Wang2021FCPjournal} in the STFT domain.

There are earlier studies using DNNs to perform full-band and sub-band modeling~\cite{Zhou2022ShorteningTarget, Quan2022NarrowbandConformer, Hao2021FullSubNet, Chen2022FullSubBandplus}.
Some differences include: (1) they only perform sub-band modelling without full-band modelling~\cite{Zhou2022ShorteningTarget, Quan2022NarrowbandConformer}; and (2) they perform sub-band modeling followed by full-band modelling~\cite{Hao2021FullSubNet, Chen2022FullSubBandplus} but without iterative information flow from sub- to full-band modules and from full- to sub-band, while we stack multiple TF-GridNet blocks to enable such an information flow so that full- and sub-band modelling can be integrated.

There are earlier studies~\cite{Yang2022TFPSNet, Xu2018GridLSTM} using LSTMs to model spectrograms along time and frequency in monaural anechoic speaker separation.
However, they do not reach very strong performance.

\subsection{Cross-Frame Self-Attention Module}

In the cross-frame self-attention module (shown in Fig.~\ref{GridNet_overview}), we first compute frame embeddings at each frame using the T-F embeddings within that frame, and then use full-utterance self-attention on these frame embeddings to model long-range context information.
The motivation is that the information flow between two T-F units needs to go through many steps in the intra-frame full-band and sub-band temporal BLSTMs, and the self-attention module enables each frame to directly attend to any frames of interest to allow for more direct information flow.
We follow the self-attention mechanism proposed in~\cite{Liu2020Attn, Pandey2021}, which is designed for U-Net based monaural music source separation and speech denoising.
\ZQHL{In contrast}, we use multi-head attention instead of single-head and we use the self-attention mechanism with \ZQHL{the proposed sub-band and full-band modules} rather than with U-Net for single- and multi-microphone speech separation.

The self-attention module has $L$ heads.
In each head $l$, we apply point-wise Conv2D, PReLU, LN along the channel and frequency dimensions (denoted as cfLN), and reshape layers to respectively obtain 2D query $Q_l \in \RR^{T\times (F\times E)}$, key $K_l \in \RR^{T\times (F\times E)}$ and value $V_l \in \RR^{T\times (F\times D/L)}$ tensors. %
The point-wise Conv2D layers for computing the query and key tensors have $E$ output channels, leading to $F \times E$-dimensional query and key vectors at each frame. %
Similarly, the point-wise Conv2D layer for computing the value tensor has $D/L$ output channels, leading to an $F \times D/L$-dimensional value vector at each frame.
All the three point-wise Conv2D layers has $D$ input channels.
Following~\cite{Vaswani2017}, we compute the attention output $A_l \in \RR^{T\times (F\times D/L)}$ by:
\begin{align}%
A_l = \text{softmax}(\frac{Q_l K_l^{\T}}{\sqrt{F\times E}})V_l.
\end{align}
We then concatenate the attention outputs of all the $L$ heads along the second dimension, reshape it back to $D\times T\times F$, apply a point-wise Conv2D with $D$ input and $D$ output channels followed by a PReLU and a cfLN to aggregate cross-head information.
Next, we add it to the input tensor $Z_b$ via a residual connection to obtain the output tensor $R_{b+1}$, which is fed to the next TF-GridNet block.

This self-attention mechanism only adds a negligible number of parameters by using point-wise Conv2D layers.
It operates at the frame level and the memory cost on attention matrices is $\mathcal{O}(B\times L \times T^2)$.
In comparison, TFPSNet~\cite{Yang2022TFPSNet} uses multi-head self-attention in each path-scanning module, and the memory cost on attention matrices is $\mathcal{O}\big(B\times L \times F \times T^2\big) + \mathcal{O}\big(B\times L \times T \times F^2\big)$, which is much higher.

\subsection{Loss Functions}\label{loss}

Since evaluation metrics usually change with datasets, we use different loss functions for different datasets, considering that different loss functions have their strengths and weaknesses~\cite{Wang2021compensation}. %
This section describes two loss functions, SI-SDR and Wav+Mag, both defined based on the re-synthesized signals of predicted RI components.
They have been proposed in earlier studies.
Our novelty is a mixture-constraint loss term to be used with SI-SDR and Wav+Mag.

\subsubsection{SI-SDR Loss with Mixture Constraint}\label{si_sdr_loss_description}

For anechoic speaker separation, there is only DNN$_1$, without the linear-filtering module and DNN$_2$. 
The model in this case is trained with utterance-level PIT~\cite{Kolbak2017}.
The loss function follows the SI-SDR loss~\cite{LeRoux2019, Luo2019}, but with two differences.

First, in the original SI-SDR metric paper~\cite{LeRoux2019}, there are two definitions for SI-SDR.
One scales \textit{source} to equalize its gain with that of estimate, and the other instead scales \textit{estimate}.
The SI-SDR loss proposed in the seminal DANet~\cite{Chen2017a} and Conv-TasNet~\cite{Luo2019} studies (and almost all the follow-up studies) uses the former, while our study uses the latter:
\begin{align}\label{sinrloss}
\mathcal{L}_{\text{SI-SDR-SE}} = - \sum\nolimits_{c=1}^{C} 10\,\text{log}_{10} \frac{\| s_q^{(c)} \|_2^2}{\| \hat{\alpha}_q^{(c)}\hat{s}_q^{(c)} - s_q^{(c)} \|_2^2},
\end{align}
where $\| \cdot \|_2^2$ computes the $L_2$ norm, $\hat{s}_q^{(c)}$ is the re-synthesized signal based on the predicted RI components for speaker $c$, $\hat{\alpha}_q^{(c)}={{\text{argmin}}}_{\alpha}\,\| \alpha \hat{s}_q^{(c)} - s_q^{(c)} \|_2^2=(\hat{s}_q^{(c)})^{\T}s_q^{(c)}/(\hat{s}_q^{(c)})^{\T}\hat{s}_q^{(c)}$, and the ``SE'' in $\mathcal{L}_{\text{SI-SDR-SE}}$ means ``scaling estimate''.
We observe that this loss leads to similar performance and faster convergence, compared with the former.

Second, we add a loss term between the summation of target sources and that of scaled estimated sources:
\begin{align}\label{sinrloss+mc}
\mathcal{L}_{\text{SI-SDR-SE+MC}} = &\mathcal{L}_{\text{SI-SDR-SE}}\,\,+ \nonumber \\
&\frac{1}{N} \Big\| \sum\nolimits_{c=1}^{C} \hat{\alpha}_q^{(c)}\hat{s}_q^{(c)} - \sum\nolimits_{c=1}^{C} s_q^{(c)} \Big\|_1,
\end{align}
where $\| \cdot \|_1$ computes the $L_1$ norm \ZQHL{and $N$ denotes the number of samples}.
Since $y_q=\sum_{c=1}^{C} s_q^{(c)}$ in our considered task of monaural, anechoic speaker separation, we name the loss term as mixture-constraint (MC) loss.
It is motivated by a trigonometric perspective~\cite{Wang2019Trigonometric} in source separation, which suggested that constraining the separated sources to sum up to the mixture yields better phase estimation.
We point out that $\sum_{c=1}^{C} \hat{\alpha}_q^{(c)}\hat{s}_q^{(c)}$ would not equal $y_q$ at run time.
This distinguishes our loss from mixture consistency~\cite{Wisdom2018MixtureConsistency}, which enforces the separated sources to sum up to the mixture.
Our loss is also different from another mixture consistency loss proposed in~\cite{Zmolikova2021}, where the DNN is trained for real-valued phase-sensitive masking without phase estimation and the task is target speaker extraction based meeting transcription.

\ZQHL{
In Eq. (\ref{sinrloss+mc}), we do not include a weighting term between the two losses for two reasons.
First, this can avoid a weighting term to tune.
Second, nowadays it is common for speaker separation systems to obtain more than $10$ dB SI-SDRi, and when the sample variance of the input mixture has been normalized to $1.0$ (which is the case in our study), the second term in our experiments has a scale less than $0.01$ when the models converge.
This way, the first term dominates the combined loss. This is desirable as the first term is directly related to the final separation performance.
}

\subsubsection{Wav+Mag Loss}\label{wav_mag_loss}

Following~\cite{Wang2021compensation}, we define the loss on the re-synthesized signal and its magnitude:
\begin{align}\label{wav+mag}
&\mathcal{L}_{\text{Wav+Mag}} =  
\sum\nolimits_{c=1}^C \Big( \frac{1}{N} \| \hat{s}_q(c) - s_q(c) \|_1 + \nonumber \\
&\,\,\,\,\,\,\,\,\,\,\,\,\,\,\,\,\,\,\,\,\,\,\,\,\frac{1}{T\times F} \Big\| \Big|\text{STFT}(\hat{s}_q(c))\Big| - \Big|\text{STFT}(s_q(c))\Big| \Big\|_1 \Big),
\end{align}
where $|\cdot|$ computes magnitude and $\text{STFT}(\cdot)$ extracts a complex spectrogram.
It has been demonstrated in~\cite{Wang2021compensation} that the magnitude loss can improve metrics such as perceptual evaluation of speech quality (PESQ), short-time objective intelligibility~\cite{H.Taal2011} (STOI), and word error rates (WER) which favor signals with a good magnitude, at a degradation on time-domain metrics such as SI-SDR.
When $C>1$, we can also add a mixture-constraint loss, similarly to Eq.~(\ref{sinrloss+mc}):
\begin{align}\label{wav+mag+MC}
&\mathcal{L}_{\text{Wav+Mag+MC}} =  
\sum\nolimits_{c=1}^C \Big( \frac{1}{N} \| \hat{s}_q(c) - s_q(c) \|_1 + \nonumber \\
&\,\,\,\,\,\,\,\frac{1}{T\times F} \Big\| \Big|\text{STFT}(\hat{s}_q(c))\Big| - \Big|\text{STFT}(s_q(c))\Big| \Big\|_1\Big) + \nonumber \\
&\,\,\,\,\,\,\,\,\,\,\,\,\,\,\,\,\,\,\frac{1}{N}\Big\| \sum\nolimits_{c=1}^C \hat{s}_q(c) - \sum\nolimits_{c=1}^C s_q(c) \Big\|_1 + \nonumber \\
&\,\frac{1}{T\times F} \Big\| \Big|\text{STFT}\Big(\sum_{c=1}^{C} \hat{s}_q(c)\Big)\Big| - \Big|\text{STFT}\Big(\sum_{c=1}^{C} s_q(c)\Big)\Big| \Big\|_1.
\end{align}

\ZQHL{In Eq.~(\ref{wav+mag}) and (\ref{wav+mag+MC}), we do not use a weighting term, as the time-domain loss and the frequency-domain loss are on a similar scale due to the Parseval's theorem.}

\section{Beamforming and Sub-Band Modelling}\label{sec:MCMFWF}

This section proposes a novel DNN-supported beamformer and connects it with integrated sub- and full-band modeling.

\subsection{\ZQHL{DNN-Supported} Multi-Frame Wiener Filter}\label{beamforming_discussion}

Assuming that target speakers are non-moving within each utterance and based on the estimated target speech $\hat{S}_q^{(1)}(c)$ by DNN$_1$, we compute a time-invariant MFWF per frequency by solving the minimization problem below:
\begin{align}\label{MCMFWF}
\underset{\mathbf{w}_q(c,f)}{{\text{argmin}}} 
\sum\nolimits_{t=1}^T \big|
\hat{S}_q^{(1)}(c,t,f) - \mathbf{w}_q(c,f)^{\H} \widetilde{\mathbf{Y}}(t,f)
\big|^2,
\end{align}
where $\widetilde{\mathbf{Y}}(t,f)=[\mathbf{Y}(t-\Delta_l,f)^\T,\dots,\mathbf{Y}(t,f)^\T,\dots,\mathbf{Y}(t+\Delta_r,f)^\T]^\T$ stacks the mixtures at nearby T-F units, $\mathbf{w}_q(c,f)\in \CC^{(\Delta_l+1+\Delta_r)\times P}$ is a time-invariant linear filter, and $(\cdot)^{\H}$ computes complex Hermitian.
$\Delta_l$ ($\geq 0$) and $\Delta_r$ ($\geq 0$) control the context of frames for filtering, resulting in a single-frame Wiener filter when $\Delta_l$ and $\Delta_r$ are both zeros and an MFWF otherwise.
A closed-form solution is available:
\begin{align}\label{mcwfcov}
&\hat{\mathbf{w}}_q(c,f) \nonumber \\
&= \Big( \sum_{t=1}^T \widetilde{\mathbf{Y}}(t,f) \widetilde{\mathbf{Y}}(t,f)^{\H} \Big)^{-1} \sum_{t=1}^T \widetilde{\mathbf{Y}}(t,f) \Big(\hat{S}_q^{(1)}(c,t,f)\Big)^{*},
\end{align}
where $(\cdot)^{*}$ computes complex conjugate.
The filtering result $\hat{S}_q^{\text{MFWF}}(c)$ is computed as
\begin{equation}\label{}
 \hat{S}_q^{\text{MFWF}}(c,t,f) = \hat{\mathbf{w}}_q(c,f)^{\H} \widetilde{\mathbf{Y}}(t,f).
\end{equation}
We name MFWF as MCMFWF when $P>1$ and as single-channel MFWF (SCMFWF) when $P=1$.

The idea of MCMFWF was proposed in~\cite{Wang2021seq}.
Differently, we use multi-microphone complex spectral mapping to obtain $\hat{S}_q^{(1)}(c)$, which consists of DNN-estimated magnitude and phase, while the system in~\cite{Wang2021seq}, even in multi-microphone cases, performs monaural, real-valued magnitude masking to obtain $\hat{S}_q^{(1)}(c)$, which consists of DNN-estimated magnitude and the mixture phase. 
It should be noted that in our recent studies~\cite{Wang2021LowDistortion, Wang2022LowLatency}, we proposed to project the mixture to DNN-estimated target speech using Eq.~(\ref{MCMFWF}), but the beamformer is single-frame (i.e., $\Delta_l=0$ and $\Delta_r=0$).
We will show in our experiments that single-frame filtering leads to worse performance than multi-frame filtering, likely due to its insufficient degrees of freedom for suppressing non-target signals.

In monaural conditions, Eq.~(\ref{mcwfcov}) becomes an SCMFWF, which can reduce reverberation by exploiting the correlations among nearby frames due to reverberation.
It is similar to the inverse convolutive prediction filter proposed in~\cite{Wang2021FCPjournal}.
The key different is that, in~\cite{Wang2021FCPjournal}, only past frames are filtered (i.e., $\Delta_l>0$ and $\Delta_r=0$).
However, future frames are also correlated with the current frame and they can also be linear-filtered to reduce the reverberation at the current frame.

In the literature, convolutional beamformer~\cite{Nakatani2019ConvBeamformer} and WPE~\cite{Nakatani2010} are the most popular multi-frame linear filters.
In their DNN-supported versions, DNN-estimated target magnitude is used in a maximum-likelihood objective for filter computation~\cite{Kinoshita2017}.
We will show in our experiments that the output of the proposed MFWF improves the performance of DNN$_2$ by a larger factor.

\subsection{Discussion on Beamforming and Sub-Band Modelling}\label{beamforming_discussion}

When beamforming results are used as extra features for DNN training (e.g., in the way shown in Fig.~\ref{system_overview}), large improvement has been observed in earlier studies~\cite{Wang2020css, Wang2021LowDistortion} (see for example the last two rows of Table~\ref{dereverb_8ch_results}).
One interesting observation is that the DNNs in these studies usually perform full-band modelling, where one typical approach is to use an encoder to encode each frame into an embedding, perform sequence modelling to refine the frame embeddings, and use a decoder to reconstruct target speech from the refined embeddings.
The encoder, for example, can be just a linear fully-connected layer followed by a non-linear activation~\cite{Luo2019} or contain a stack of non-linear layers in the form of a UNet-style encoder~\cite{Wang2020css}.
Our insight is that the large improvement is likely because the beamformers are computed based on signals only within each sub-band and the beamforming results could hence be complementary to full-band modeling, which simultaneously models all the frequencies but may not be good at sufficiently modeling each frequency since different frequencies exhibit diverse spectral, temporal and spatial patterns (see also our discussions in Section~\ref{fullsubband_discussion}).

Each sub-band temporal module in TF-GridNet models each frequency using a BLSTM shared across all the frequencies \ZQHL{to mimic sub-band filtering}.
This could be a better way of \textit{neural beamforming} than earlier approaches where DNNs are mainly used for full-band modeling.
In our best-performing system, we still compute an MCMFWF result based on the output of a first TF-GridNet and use a second TF-GridNet for post-filtering (i.e., Fig.~\ref{system_overview}).
This can be viewed as another way of full- and sub-band integrated modeling, and is found to improve the performance of using just one single TF-GridNet, but the improvement brought by the beamformer \ZQHL{followed by post-filtering} is much less impressive than the one achieved when the two DNNs are trained to perform full-band modelling.
See also our discussion later in Section~\ref{result_dereverb}.

We point out that the sub-band (\textit{a.k.a} narrow-band) property for per-frequency modeling is afforded by STFT.
This property bears an important advantage of STFT-domain approaches: we can exploit intra- and cross-frequency information to achieve potentially better separation.
In comparison, the learned bases by time-domain models are usually not narrow-band~\cite{Luo2019, Cornell2022FB}, and many current time-domain models do not have a concept of sub-band or narrow-band frequency to exploit.%

\section{Experimental Setup}\label{experiments}

We evaluate the proposed algorithms on five tasks, including speaker separation in anechoic, reverberant and noisy-reverberant conditions, speech dereverberation, and noisy-reverberant speech enhancement.
This section describes the setup for each task, baselines, and miscellaneous configurations.
Our experiments cover major speech separation tasks and we use public datasets with existing published results to highlight that the improvements obtained in our study are relative to very strong baselines.

\subsection{Setup for Monaural, Anechoic Speaker Separation}\label{wsj02mix_setup}

We use \textbf{WSJ0-2mix}~\cite{Hershey2016}, the most popular dataset to benchmark monaural talker-independent speaker separation algorithms in anechoic conditions.
It has 20,000 ($\sim$30.4 h), 5,000 ($\sim$7.7 h) and 3,000 ($\sim$4.8 h) two-speaker mixtures respectively in its training, validation and test sets.
The clean source signals are sampled from the WSJ0 corpus.
The speakers in the training and validation sets are different from the speakers for testing.
\ZQHL{The two utterances in each of the mixtures available in WSJ0-2mix are fully-overlapped, and their relative energy level is uniformly sampled from the range $[-5, 5]$ dB when WSJ0-2mix is created.}
The sampling rate is 8 kHz.

\subsection{Setup for Reverberant Speaker Separation}\label{smswsj_setup}

We use \textbf{SMS-WSJ}~\cite{Drude2019}, a popular corpus for comparing two-speaker separation algorithms in reverberant conditions.
The clean speech is sampled from the WSJ0 and WSJ1 datasets.
The corpus contains 33,561 ($\sim$87.4 h), 982 ($\sim$2.5 h) and 1,332 ($\sim$3.4 h) two-speaker mixtures for training, validation and testing, respectively.
The simulated microphone array has six microphones arranged uniformly on a circle with a diameter of 20 cm.
For each mixture, the speaker-to-array distance is drawn from the range $[1.0, 2.0]$ m, and T60 from $[0.2, 0.5]$ s.
A weak white noise is added to simulate microphone sensor noises, and the energy level between the sum of the reverberant speech signals and the noise is sampled from the range $[20, 30]$ dB.
The sampling rate is 8 kHz.

For ASR evaluation, the default Kaldi-based ASR backend provided with SMS-WSJ~\cite{Drude2019} is used.
It is trained using single-speaker noisy-reverberant speech as inputs and the state alignments of its corresponding direct-path signal as labels.
A standard tri-gram language model is used for decoding.

We perform joint denoising, dereverberation and separation.
We consider one-, two- and six-channel tasks, and use the direct-path signals as the training target.
For two-channel processing, we take the signals at microphone 1 and 4 as input, and for monaural separation, we use the signal at microphone 1.
The first microphone is always used as the reference.

\subsection{Setup for Noisy-Reverberant Speaker Separation}\label{whamr_setup}

We use \textbf{WHAMR!}~\cite{Maciejewski2020} to validate our algorithms for noisy-reverberant speaker separation.
It re-uses the two-speaker mixtures in WSJ0-2mix~\cite{Hershey2016} but reverberates each clean source and adds non-stationary noises. %
In each mixture, the T60 is sampled from the range $[0.2, 1.0]$ s, signal-to-noise ratio (SNR) between the louder speaker and noise from $[-6, 3]$ dB, relative energy level between the two speakers from $[-5, 5]$ dB, and speaker-to-array distance from $[0.66, 2.0]$ m.
There are 20,000 ($\sim$30.4 h), 5,000 ($\sim$7.7 h) and 3,000 ($\sim$4.8 h) binaural mixtures respectively for training, validation and testing.
We use its \textit{min} and 8 kHz version.

We aim at joint dereverberation, denoising and speaker separation.
The direct-path signal of each speaker at the first microphone is used as the target for training and as the reference for metric computation.

\subsection{Setup for Speech Dereverberation}\label{reverbwsj0cam_setup}

We use a simulated reverberant dataset with weak air-conditioning noises, since there lacks a well-designed popular dataset for speech dereverberation.\footnote{We considered the REVERB corpus~\cite{Kinoshita2016}, but its training set is simulated based on 24 eight-channel RIRs, which are too few for training DNN models.}
Although simulated by ourselves, this dataset has been used in our recent studies~\cite{Wang2021FCPjournal, Wang2021LowDistortion}, which reported very strong results.
The clean source signals for simulation are from the WSJCAM0 corpus, which includes 7,861, 742 and 1,088 utterances respectively in its training, validation and test sets.
Based on them, we simulate 39,293 ($\sim$77.7 h), 2,968 ($\sim$5.6 h), and 3,262 ($\sim$6.4 h) noisy-reverberant mixtures respectively as our training, validation, and test sets.
The data spatialization process follows~\cite{Wang2020dMCCSMconference}, where, for each utterance, we randomly sample a room with random room characteristics and speaker and microphone locations, using the Pyroomacoustics RIR generator~\cite{Scheibler2018}.
The simulated microphone array has eight microphones arranged on a circle with a diameter of 20 cm.
The speaker-to-array distance is drawn from the range $[0.75, 2.5]$ m and T60 from $[0.2, 1.3]$ s.
For each utterance, an eight-channel diffuse air-conditioning noise is sampled from the REVERB dataset~\cite{Kinoshita2016} and added to the reverberant speech, and the SNR between the direct-path signal and the noise is sampled from the range $[5, 25]$ dB.
The sampling rate is 16 kHz.
We denote this dataset as \textbf{WSJ0CAM-DEREVERB}.

We aim at removing any early reflections and late reverberation.
The direct-path signal of the target speaker at the first microphone is used as %
the reference for metric computation.

\subsection{Setup for Noisy-Reverberant Speech Enhancement}\label{l3das_setup}

The \textbf{L3DAS22} 3D speech enhancement task~\cite{Guizzo2022L3DAS} challenges participants to reconstruct the dry speech source signal from its far-field mixture simulated by using two four-channel Ambisonic-format signals in a noisy-reverberant office environment.
The dry source signals are drawn from LibriSpeech and noise signals from FSD50k~\cite{fonseca2020fsd50k}.
The SNR is sampled from the range $[6, 16]$ dB. %
Real RIRs are used for simulation. Such RIRs were recorded in an office room by using two first-order A-format Ambisonic arrays, each with four microphones.
The microphone placement is fixed, with one Ambisonic microphone array placed at the room center and the other being 20\,cm away.
The room configuration is the same between training and testing, and the source positions are sampled uniformly inside the room with no overlap of positions between training and testing.
Artificial mixtures are generated by convolving dry speech and dry noise signals with the measured RIRs and the convolved signals are then added together.
There are 37,398 ($\sim$81.3 h), 2,362 ($\sim$3.9 h) and 2,189 ($\sim$3.5 h) mixtures respectively in the training, validation and test sets.
The generated A-format Ambisonic mixtures are converted to B-format Ambisonic via a transformation consisting of a pre-filter, a mixing matrix and a post-filter.
The task is to predict the dry speech based on the B-format Ambisonic mixture.
The sampling rate is 16 kHz.

The submitted systems were ranked by using a combination of STOI and WER:
\begin{equation}\label{eq:task1_metric}
     \text{Task1Metric} = \Big(\text{STOI} + (1 - \text{WER})\Big)/2.
\end{equation}
Since STOI and WER scores are both in the range of $[0,1]$, the composite metric is also in $[0,1]$.
The WER is computed from the transcription of enhanced speech with that of the dry speech, both decoded by a pre-trained wav2vec2 ASR model.

Differently from the other setups, the goal in this task is to predict the dry speech from far-field multi-channel mixtures.
This requires the submitted systems to not only remove reverberation and noises, but also to time-align the estimated speech with the dry speech (as STOI degrades with misalignment), which requires the systems to perform implicit or explicit localization of the target source so that a time-aligned estimate can be obtained. 
This is achievable since the Ambisonic arrays form a fixed three-dimensional geometry.

\subsection{Baselines}

We can compare our approaches with others by using system-level performance.
For MFWF, we provide the results of other linear filters, including (1) in multi-channel cases, convolutional beamformer~\cite{Nakatani2019ConvBeamformer}; and (2) in monaural cases, WPE~\cite{Nakatani2010, Kinoshita2017}.
We replace the MFWF module between DNN$_1$ and DNN$_2$ in Fig.~\ref{system_overview} with a DNN-supported convolutional beamformer or WPE filter to compare their effectiveness at improving DNN$_2$. %

\subsubsection{System-Level Baselines}

Since the datasets in all the considered tasks have existing results reported in earlier studies, we can compare our results with the strongest ones achieved by competing approaches.
Notably, we will compare with our previous studies~\cite{Wang2020css, Wang2021FCPjournal, Wang2021FCPwaspaa, Lu2022}, which also follow the MISO-BF-MISO approach shown in Fig.~\ref{system_overview} but uses TCN-DenseUNet and other sub-band linear filters.

\subsubsection{Baseline for MCMFWF}

In multi-channel cases, we consider convolutional beamformer~\cite{Nakatani2019ConvBeamformer}, a very popular multi-channel multi-frame filter in speech separation, as the baseline.
We compute it by solving the problem~\cite{Nakatani2019ConvBeamformer} below:
\begin{align}\label{convbf_objective}
\underset{\mathbf{w}_q(c,f)}{{\text{argmin}}} \sum\nolimits_{t=1}^{T} \frac{|\mathbf{w}_q(c,f)^{\H}\ \Bar{\mathbf{Y}}(t,f)|^2}{\hat{\lambda}_q(c, t, f)} \nonumber \\ \text{subject to}\,\,\,\,\mathbf{w}_{q;0}(c,f)^{\H}\hat{\mathbf{d}}_q(c,f) = 1,
\end{align}
where $\Bar{\mathbf{Y}}(t,f)=[\mathbf{Y}(t-\Delta_d-\Delta_l+1,f)^\T,\dots,\mathbf{Y}(t-\Delta_d,f)^\T, \mathbf{Y}(t,f)^\T]^\T \in \CC^{(\Delta_l+1)\times P}$ with $\Delta_d$ denoting a prediction delay and $\Delta_l$ the number of filter taps for past frames beyond the prediction delay, $\mathbf{w}_q(c,f)=[\mathbf{w}_{q;-\Delta_d-\Delta_l+1}(c,f)^\T,\dots,\mathbf{w}_{q;-\Delta_d}(c,f)^\T,\mathbf{w}_{q;0}(c,f)^\T]^\T \in \CC^{(\Delta_l+1)\times P}$ with $\mathbf{w}_{q;i}(c,f)\in \CC^P$ denoting the filter applied to frame $t+i$ in order to produce the result at the current frame $t$, and $\hat{\mathbf{d}}_q(c,f)$ is the estimated relative transfer function for microphone $q$.
Following~\cite{Drude2020NARAWPE} and based on the DNN-estimated target speech $\hat{S}_q^{(1)}(c)$, $\hat{\lambda}_q(c)$, the estimated power spectral density of target speech, can be computed as:
\begin{align}\label{wpelambda}
\hat{\lambda}_q(c,t,f)=\text{max}\Big(\varepsilon\,\, \text{max}(|\hat{S}_q^{(1)}(c)|^2),|\hat{S}_q^{(1)}(c,t,f)|^2\Big),
\end{align}
where $\text{max}(\cdot)$ extracts the maximum value of a spectrogram, $\text{max}(\cdot,\cdot)$ returns the larger of two values, and $\varepsilon$ is a floor value to avoid putting too much weight on T-F units with low energy.
Through T-F masking and also based on the DNN-estimated target speech $\hat{S}_q^{(1)}(c)$, $\hat{\mathbf{d}}_q(c,f)$ is computed as the principal eigenvector of an estimated speech covariance matrix~\cite{Yoshioka2015, Heymann2015, Gannot2017} for non-moving point sources, i.e.,
\begin{align}
\hat{\mathbf{\Phi}}(c,f) &= \sum\nolimits_{t=1}^T \hat{m}(c,t,f) \mathbf{Y}(t,f)\mathbf{Y}(t,f)^{\H}, \\
\hat{m}(c,t,f) &= \frac{|\hat{S}_q^{(1)}(c,t,f)|}{|\hat{S}_q^{(1)}(c,t,f)| + |Y_q(t,f) - \hat{S}_q^{(1)}(c,t,f)|},
\end{align}
\begin{align}
\hat{\mathbf{d}}(c,f) &= \mathcal{P}\big(\hat{\mathbf{\Phi}}(c,f)\big), \\
\hat{\mathbf{d}}_q(c,f) &= \hat{\mathbf{d}}(c,f) / \hat{d}(c,f;q),
\end{align}
where $\mathcal{P}(\cdot)$ extracts the principal eigenvector, and $\hat{d}(c,f;q)$ denotes the $q^{\text{th}}$ element in $\hat{\mathbf{d}}(c,f)$.
The results of convolutional beamformer is comptued as
\begin{align}\label{convbf_result}
\hat{S}_q^{\text{ConvBF}}(c,t,f) = \hat{\mathbf{w}}_q(c,f)^{\H}\ \Bar{\mathbf{Y}}(t,f),
\end{align}
where ``ConvBF'' denotes convolutional beamformer.

Notice that our DNN-supported MCMFWF in Eq.~(\ref{MCMFWF}) is simpler to compute than the convolutional beamformer.

\subsubsection{Baseline for SCMFWF}

In the single-microphone case, convolutional beamformer turns into the WPE filter~\cite{Nakatani2010}.
Following the DNN-WPE algorithm~\cite{Kinoshita2017}, we compute it by using the magnitude of $\hat{S}_q^{(1)}(c)$ estimated by DNN$_1$.
The filter is computed by solving the following problem:
\begin{align}\label{dnn_wpe}
\underset{\mathbf{w}_q(c,f)}{{\text{argmin}}} \sum\nolimits_{t=1}^{T} \frac{|Y_q(t,f) - \mathbf{w}_q(c,f)^{\H}\ \Breve{\mathbf{Y}}(t-\Delta_d,f)|^2}{\hat{\lambda}_q(c, t, f)},
\end{align}
where $\Breve{\mathbf{Y}}_q(t,f)=[Y_q(t-\Delta_l+1,f)^\T,\dots,Y_q(t,f)^\T]^\T \in \CC^{\Delta_l}$, $\mathbf{w}_q(c,f)\in \CC^{\Delta_l}$, $\Delta_d$ is a prediction delay, and $\hat{\lambda}_q(c)$ is computed using Eq.~(\ref{wpelambda}).
The WPE result is obtained as
\begin{align}\label{wpe_result}
\hat{S}_q^{\text{WPE}}(c,t,f)=Y_q(t,f) - \hat{\mathbf{w}}_q(c,f)^{\H} \Breve{\mathbf{Y}}(t-\Delta_d,f).
\end{align}

\subsection{Miscellaneous Setup}

In default, for STFT, the window length is $32$ ms and hop length $8$ ms, and the square-root Hann window is used as the analysis window.
In this case, for $16$ kHz sampling rate, a $512$-point discrete Fourier transform (DFT) is applied to extract $257$-dimensional complex STFT spectra at each frame, and for $8$ kHz, a $256$-point DFT is used to extract $129$-dimensional complex STFT spectra.
$E$ (see its definition in Table~\ref{summary_hyperparam}) is set to $4$ for 8 kHz and to $2$ for $16$ kHz.
This way, the dimension of frame-level embeddings (i.e., $F\times E$) used for self-attention is reasonable.

For MFWF, we set $\Delta_l$ and $\Delta_r$, which controls the filter taps, to $4$ and $3$ for eight-channel separation,
to $5$ and $4$ for six-channel, to $15$ and $14$ for two-channel, and to $20$ and $19$ for single-channel.
For convolutional beamformer, we set the prediction delay $\Delta_d$ to 3 following~\cite{Nakatani2019ConvBeamformer}, and tune $\Delta_l$ to $7$ for eight-channel processing, to $9$ for six-channel, and to $29$ for two-channel.
For WPE, $\Delta_d$ is also 3 and $\Delta_l$ is tuned to $40$.
We emphasize that a positive prediction delay $\Delta_d$ is found important for convolutional beamformer and WPE to avoid target cancellation~\cite{Nakatani2010, Nakatani2019ConvBeamformer}, and both filters are designed by the original authors to not filter future frames, because future frames contain the reverberation of the target speech at the current frame and including them for linear filtering would lead to target cancellation.
$\varepsilon$ in Eq.~(\ref{wpelambda}) is tuned to $10^{-5}$.

In each epoch, we sample a $4$-second segment from each mixture for training.
We normalize the sample variance of each mixture segment to $1.0$ and use the same scaling factor to scale the target sources, before using them for training.
Adam is used as the optimizer.
The $L_2$ norm for gradient clipping is set to $1.0$.
The learning rate starts from $0.001$ and is reduced by half if the validation loss does not improve in $3$ epochs.

We do not use any dynamic mixing or data augmentation~\cite{Subakan2021}.

\begin{table*}[t]
\scriptsize
\centering
\sisetup{table-format=2.2,round-mode=places,round-precision=2,table-number-alignment = center,detect-weight=true,detect-inline-weight=math}
\caption{\textsc{Masking and Mapping Comparison Based on WSJ0-2mix.}}
\vspace{-0.1cm}
\label{result_fair}
\setlength{\tabcolsep}{2pt}
\begin{tabular}{
c
c
c
c
c %
c %
S[table-format=3,round-precision=0]
S[table-format=1,round-precision=0]
S[table-format=1,round-precision=0]
S[table-format=3,round-precision=0]
S[table-format=1.1,round-precision=1] %
S[table-format=2.1,round-precision=1] %
c %
S[table-format=2.1,round-precision=1] %
}
\toprule

{\multirow{2}{*}{\rotatebox[origin=c]{0}{Row}}}  & {\multirow{2}{*}{Systems}} & {Unfold+LN or} & {Use} & {\multirow{2}{*}{Masking or Mapping?}} & {Window/hop} & {\multirow{2}{*}{$D$}} & {\multirow{2}{*}{$I$}} & {\multirow{2}{*}{$J$}} & {\multirow{2}{*}{$H$}} & {\#params} & {\multirow{2}{*}{GMAC/s}} & {\multirow{2}{*}{Loss}} & {SI-SDRi} \\
& & {LN+Unfold} & {attention?} & & {sizes (ms)} & & & & & {(M)} & & & {(dB)} \\

\midrule

1 & DPRNN~\cite{Luo2020} & {-} & {-} & Embedding masking & {$0.25/0.125$} & {-} & {-} & {-} & {-} & 2.6 & 42.2025  & (\ref{sinrloss}) & 18.8 \\

\midrule

2 & TF-GridNet & {Unfold+LN} & \xmark & Embedding masking & {$32/8$} & 64 & 1 & 1 & 128 & 2.8 & 47.551632551 & (\ref{sinrloss}) & 20.7 \\
3 & TF-GridNet & {Unfold+LN} & \xmark & Complex ratio masking & {$32/8$} & 64 & 1 & 1 & 128 & 2.6 & 42.431269607 & (\ref{sinrloss}) & 20.8 \\
4 & TF-GridNet & {Unfold+LN} & \xmark & Complex spectral mapping & {$32/8$} & 64 & 1 & 1 & 128 & 2.6 & 42.431269607 & (\ref{sinrloss}) &\bfseries 21.2 \\

\bottomrule
\end{tabular}
\vspace{-0.2cm}
\end{table*}

\subsection{Evaluation Metrics}

The evaluation metrics vary with tasks.
We consider SI-SDR or SI-SDRi~\cite{LeRoux2019}, SDR or SDR improvement (SDRi)~\cite{Vincent2006a}, PESQ, STOI or extended STOI (eSTOI)~\cite{H.Taal2011}\footnote{\url{https://github.com/mpariente/pystoi}, v0.3.3}, and WER.
For PESQ, we use the \textit{python-pesq} toolkit\footnote{\url{https://github.com/ludlows/python-pesq}, v0.0.2} to report narrow-band MOS-LQO scores.
\ZQHL{SI-SDR and SDR measure the quality of predictions at the sample level, PESQ and STOI are objective metrics of speech quality and intelligibility respectively, and WER is a widely-used metric for measuring speech recognition performance.}

The number of model parameters is reported in millions (M). \ZQHL{We use the \textit{flops-counter.pytorch} toolkit\footnote{\ZQHL{\url{https://github.com/sovrasov/flops-counter.pytorch}.
Note that, in default, \textit{flops-counter.pytorch} only tries to count the MAC operations of a list of pre-defined modules that are already available in Pytorch.
We confirm that we also count the MAC operations of our customized modules.
}} to count the number of multiply–accumulate (MAC) operations needed to process a 4-second mixture, and report it in giga-operations per second (GMAC/s).
Following \cite{Wang2023ICASSPLowLat}, we implement Deconv1D as a linear layer followed by overlap-add. This can reduce the number of MAC operations when the stride $J$ is larger than 1, and speed up training and inference when the kernel size $I$ equals $J$ ($>1$).
}

\section{Evaluation Results}\label{results}

\ZQHL{We first show the effectiveness of TF-GridNet at separation on various tasks and datasets, and then present a study on the computational requirements of different TF-GridNet configurations and their performance on WSJ0-2mix.}

\subsection{Results on WSJ0-2mix}\label{result_WSJ02mix}

We evaluate TF-GridNet on monaural, anechoic speaker separation.
SI-SDRi~\cite{LeRoux2019} and SDRi~\cite{Vincent2006a} are used as the evaluation metrics, following previous studies.
The mixture SI-SDR is $0$ dB and the mixture SDR $0.2$ dB.
We always use $B=6$ blocks for WSJ0-2mix.

\subsubsection{Comparison with DPRNN}

\ZQHL{Table~\ref{result_fair} compares the performance of TF-GridNet with DPRNN~\cite{Luo2020}.
We configure TF-GridNet to use almost the same number of parameters and almost the same amount of computation as DPRNN.
This is implemented by using BLSTMs in each model and unifying the embedding dimension (and the bottleneck dimension in the cases of DPRNN) to $64$ and the hidden dimension of the BLSTMs to $128$. %
For DPRNN, we set the window size to $2$ samples, hop size to $1$ sample, chunk size to $250$ frames, and overlap between consecutive chunks to $50\%$, following the best configuration reported in \cite{Luo2020}.
For TF-GridNet, in each block we remove the full-band self-attention module, and set $I$ and $J$ to $1$ (in this case, the order of LN and Unfold does not matter).
From row 1 and 4, we observe that TF-GridNet with complex spectral mapping obtains better results (21.2 vs. 18.8 dB).
Table~\ref{result_fair} also reports the performance of using TF-GridNet with masking in row 2 and 3.
In row 2, we mask learned embeddings, following~\cite{Luo2019,Luo2020,Yang2022TFPSNet}.
We closely follow the encoder-masking-decoding modules used in \cite{Yang2022TFPSNet}, but replace their path-scanning modules with our intra-frame full-band and sub-band temporal modules.
In row 3, we use TF-GridNet for complex ratio masking based separation~\cite{Williamson2016, Liu2019DeepCASA}.
After obtaining the output tensor of the Deconv2D module (see Fig.~\ref{GridNet_overview}), we first truncate the values in the tensor to the range $[-5,5]$ to obtain an estimated complex ratio mask and then multiply it with the mixture spectrogram for separation.
From row 2, 3 and 4, we notice that complex spectral mapping performs better (21.2 vs. 20.7 and 20.8 dB).}

\begin{table}[t]
\scriptsize
\centering
\sisetup{table-format=2.2,round-mode=places,round-precision=2,table-number-alignment =center,detect-weight=true,detect-inline-weight=math}
\caption{\textsc{Ablation Results on WSJ0-2mix.}}
\vspace{-0.1cm}
\label{result_hyperparam}
\setlength{\tabcolsep}{1.5pt}
\begin{tabular}{
r
c
c
c
S[table-format=1,round-precision=0]
S[table-format=3,round-precision=0]
S[table-format=1,round-precision=0]
S[table-format=1,round-precision=0]
S[table-format=3,round-precision=0]
S[table-format=2.1,round-precision=1]
c
S[table-format=2.1,round-precision=1]
}
\toprule

{\multirow{2}{*}{Row}} & {\multirow{2}{*}{Systems}} & {Unfold+LN or} & Use        & {\multirow{2}{*}{$L$}} & {\multirow{2}{*}{$D$}} & {\multirow{2}{*}{$I$}} & {\multirow{2}{*}{$J$}} & {\multirow{2}{*}{$H$}} & {\#params} & {\multirow{2}{*}{Loss}} & {SI-SDRi} \\
                       &                            & {LN+Unfold} & attention? &  & & & & & {(M)} & & {(dB)} \\

\midrule

1 & TF-GridNet & {Unfold+LN} & \xmark & {-} & 64 & 1 & 1 & 128 & 2.6 & (\ref{sinrloss}) & 21.2 \\
2 & TF-GridNet & {Unfold+LN} & \xmark & {-} & 16 & 4 & 1 & 128 & 2.6 & (\ref{sinrloss}) & 20.5 \\
3 & TF-GridNet & {Unfold+LN} & \xmark & {-} & 128 & 1 & 1 & 128 & 3.6 & (\ref{sinrloss}) & 21.6 \\
4 & TF-GridNet & {Unfold+LN} & \xmark & {-} & 16 & 8 & 1 & 128 & 3.6 & (\ref{sinrloss}) & 21.6 \\

\midrule

5 & TF-GridNet & {Unfold+LN} & \xmark & {-} & 16 & 8 & 1 & 128 & 3.6 & (\ref{sinrloss+mc}) & 21.8 \\ %
6 & TF-GridNet & {Unfold+LN} & \xmark & {-} & 16 & 8 & 1 & 192 & 6.5 & (\ref{sinrloss+mc}) & 21.9 \\ %
7 & TF-GridNet & {Unfold+LN} & \xmark & {-} & 24 & 8 & 1 & 192 & 8.0 & (\ref{sinrloss+mc}) & 22.5 \\ %

\midrule

8 & TF-GridNet & {Unfold+LN} & \cmark & 1 & 24 & 8 & 1 & 192 & 8.0 & (\ref{sinrloss+mc}) & 22.6 \\ %
9 & TF-GridNet & {Unfold+LN} & \cmark & 4 & 24 & 8 & 1 & 192 & 8.1 & (\ref{sinrloss+mc}) & 22.9 \\ %
10 & TF-GridNet & {Unfold+LN} & \cmark & 4 & 48 & 4 & 1 & 192 & 8.2 & (\ref{sinrloss+mc}) & 23.0 \\
11 & TF-GridNet & {LN+Unfold} & \cmark & 4 & 48 & 4 & 1 & 192 & 8.2 & (\ref{sinrloss+mc}) & 23.175252076899632 \\
12 & TF-GridNet & {LN+Unfold} & \cmark & 4 & 64 & 4 & 1 & 256 & 14.5 & (\ref{sinrloss+mc}) & \bfseries 23.50169992613451 \\

\bottomrule
\end{tabular}
\vspace{-0.2cm}
\end{table}

\subsubsection{Ablation Results with Different Hyper-Parameters}

Table~\ref{result_hyperparam} presents the ablation results of our models on WSJ0-2mix using different model hyper-parameters.
From row 1-4, we can see that, when the kernel size is sufficiently large (i.e., $I=8$), using the Unfold and Deconv1D mechanism together with a smaller embedding dimension (i.e., $D=16$) does not decreases SI-SDRi, compared with the configuration that uses a larger embedding dimension (i.e., $D=128$) but does not stack nearby T-F embeddings (i.e., $I=1$).
One benefit of using the former configuration is that the memory consumption is lower.
From row 4 and 5, we can see that the MC loss produces slightly better SI-SDRi (21.6 vs. 21.8 dB).
\ZQHL{Comparing row 7 with 5 and 6}, we notice that enlarging the model size by increasing the number of hidden units $H$ in BLSTMs and the embedding dimension $D$ produces clear improvement.
The results in row 7, 8 and 9 suggest that including the full-band self-attention module is beneficial, and using four attention heads leads to better performance than just using one (22.9 vs. 22.6 dB).
In row 10, we increase the embedding dimension to $48$ and reduce the kernel size $I$ to 4, and obtain slightly better SI-SDRi than the model in row 9 (23.0 vs. 22.9 dB).
In row 11, we use LN+Unfold rather than Unfold+LN.
This results in 0.2 dB better SI-SDRi (23.2 vs. 23.0 dB).
Further enlarging model size in row 12 produces further gains (from 23.2 to 23.5 dB).

\begin{table}[t]
\scriptsize
\centering
\sisetup{table-format=2.2,round-mode=places,round-precision=2,table-number-alignment = center,detect-weight=true,detect-inline-weight=math}
\caption{\textsc{Performance Comparison with Other Systems on WSJ0-2mix.}}
\vspace{-0.1cm}
\label{comparison_with_others}
\setlength{\tabcolsep}{2.5pt}
\begin{tabular}{
cc
S[table-format=4,round-precision=0]
S[table-format=3.1,round-precision=1]
S[table-format=2.1,round-precision=1]
S[table-format=2.1,round-precision=1]
}
\toprule
Systems & Domain & {Year} & {\#params (M)} & {SI-SDRi (dB)}  & {SDRi (dB)} \\
\midrule
DPCL++~\cite{Isik2016} & T-F & 2016 & 13.6 & 10.8 & {-} \\
uPIT-BLSTM-ST~\cite{Kolbak2017} & T-F & 2017 & 92.7 & {-} & 10.0 \\
ADANet~\cite{Chen2017a} & T-F & 2018 & 9.1 & 10.4 & 10.8 \\
WA-MISI-5~\cite{WZQe2eMISI2018} & T-F & 2018 & 32.9 & 12.6 & 13.1 \\
Sign Prediction Net~\cite{Wang2019Trigonometric} & T-F & 2019 & 56.6 & 15.3 & 15.6 \\
Conv-TasNet~\cite{Luo2019} & Time & 2019 & 5.1 & 15.3 & 15.6 \\
Deep CASA~\cite{Liu2019DeepCASA} & T-F & 2019 & 12.8 & 17.7 & 18.0 \\
Conv-TasNet-MBT~\cite{Lam2020MBT} & Time & 2020 & 8.8 & 15.6 & {-} \\
FurcaNeXt~\cite{Shi2019FurcaNeXt} & Time & 2020 & 51.4 & {-} & 18.4 \\
SUDO RM -RF~\cite{Tzinis2020} & Time & 2020 & 2.6 & 18.9 & {-} \\
DPRNN~\cite{Luo2020} & Time & 2020 & 2.6 & 18.8 & 19.0 \\
Gated DPRNN~\cite{Nachmani2020} & Time & 2020 & 7.5 & 20.1 & 20.4 \\
DPTNet~\cite{Chen2020DPTnet} & Time & 2020 & 2.7 & 20.2 & 20.6 \\
DPTCN-ATPP~\cite{Zhu2021} & Time & 2021 & 4.7 & 19.6 & 19.9 \\
SepFormer~\cite{Subakan2021} & Time & 2021 & 26.0 & 20.4 & 20.5 \\
Sandglasset~\cite{Lam2021} & Time & 2021 & 2.3 & 20.8 & 21.0 \\
Wavesplit~\cite{Zeghidour2020} & Time & 2021 & 29.0 & 21.0 & 21.2 \\
TFPSNet~\cite{Yang2022TFPSNet} & T-F & 2022 & 2.7 & 21.1 & 21.3 \\
MTDS (DPTNet)~\cite{Qian2022} & Time & 2022 & 4.0 & 21.5 & 21.7 \\
SFSRNet~\cite{Rixen2022} & Time & 2022 & 59.0 & 22.0 & 22.1 \\ 
QDPN~\cite{Rixen2022QDPN} & Time & 2022 & 200.0 & 22.1 & {-} \\ 
\midrule
TF-GridNet & T-F & 2022 & 14.5 & \bfseries 23.50169992613451 & \bfseries 23.6073187526 \\
\bottomrule
\end{tabular}
\vspace{-0.2cm}
\end{table}

\subsubsection{Comparison with Previous Models}

Table~\ref{comparison_with_others} compares the performance of our best TF-GridNet with previous models on WSJ0-2mix.
Compared with previous best such as SepFormer~\cite{Subakan2021}, SFSRNet~\cite{Rixen2022} and QDPN~\cite{Rixen2022QDPN}, our model \ZQHL{has} a modest size.
Notice that, since 2019, T-F domain models have been largely under-explored and under-represented for anechoic speaker separation, and many research efforts have been devoted to time-domain approaches.
The recent TFPSNet model~\cite{Yang2022TFPSNet} achieves a competitive SI-SDRi at 21.1 dB, but the performance still falls within the range of scores (i.e., $[20.0,22.0]$ dB SI-SDRi) that can be commonly achieved by modern time-domain models.
Our study, for the first time since 2019, unveils that complex T-F domain models, with a contemporary DNN architecture, can outperform modern time-domain models by a large margin.
\ZQHL{Later in Section \ref{cost_perf}, we will provide the computational cost of TF-GridNet.}

\subsection{Results on SMS-WSJ and WHAMR!}\label{result_smswsj}

This section evaluates TF-GridNet and the two-DNN system on reverberant and noisy-reverberant speaker separation.
In the following experiments, in default we use $B=4$ and $H=192$ to save computation\footnote{We also experimented with larger TF-GridNets and observed better performance but we consider this unnecessary. We will show later that TF-GridNet with this setup already produces better results than competing models.} and use LN+Unfold.
Based on the validation sets, we set $I=4$, $J=1$ and $D=48$ for SMS-WSJ, and $I=8$, $J=1$ and $D=24$ for WHAMR!.

\begin{table}[]
\scriptsize
\centering
\sisetup{table-format=2.2,round-mode=places,round-precision=2,table-number-alignment = center,detect-weight=true,detect-inline-weight=math}
\caption{\textsc{Results on SMS-WSJ (1ch).}}
\vspace{-0.1cm}
\label{sms_wsj_results_1ch}
\setlength{\tabcolsep}{1.75pt}
\begin{tabular}{l
c %
c %
c
S[table-format=2.1,round-precision=1]
S[table-format=2.1,round-precision=1]
S[table-format=1.2,round-precision=2]
S[table-format=1.3,round-precision=3]
S[table-format=2.2,round-precision=2]
}
\toprule
{\multirow{2}{*}{Systems}} & {\multirow{2}{*}{$\Delta_l$}} & {\multirow{2}{*}{$\Delta_r$}} & {\multirow{2}{*}{Loss}} & {SI-SDR} & {SDR} & {\multirow{2}{*}{PESQ}} & {\multirow{2}{*}{eSTOI}} & {WER} \\
 & & & & {(dB)} & {(dB)} & & & {(\%)} \\

\midrule
Unprocessed & {-} & {-} & - & -5.5 & -0.4 & 1.50 & 0.441 & 78.4 \\
\midrule
DNN$_1$ & {-} & {-} & (\ref{sinrloss+mc}) & 16.1956 & 17.249209 & 3.45168 & 0.9238977 & 9.49 \\
DNN$_1$ & {-} & {-} & (\ref{wav+mag}) & 14.72918 & 15.747810 & 3.35166746 & 0.91447677 & 9.64 \\
DNN$_1$ & {-} & {-} & (\ref{wav+mag+MC}) & 15.66463337 & 16.604234733 & 3.4118591 & 0.9239642 & 9.26 \\ DNN$_1$+DNN$_2$ & {-} & {-} & (\ref{wav+mag+MC}) & 16.6 & 17.6 & 3.53 & 0.93449799 & 8.80 \\
DNN$_1$+SCMFWF+DNN$_2$ & 39 & 0 & (\ref{wav+mag+MC}) & 17.595673 & 18.66561035 & 3.66981964 & 0.9459155409 & 8.51 \\
DNN$_1$+SCMFWF+DNN$_2$ & 20 & 19 & (\ref{wav+mag+MC}) & \bfseries 18.4490 & \bfseries 19.62540 & \bfseries 3.6988555 & \bfseries 0.95216320 & \bfseries 7.91 \\
\midrule
DNN$_1$+WPE+DNN$_2$ & 40 & {-} & (\ref{wav+mag+MC}) & 17.536703685113977 & 18.571642394523618 & 3.668501243755982 & 0.9465367005852843 & 8.19 \\
\midrule
DPRNN-TasNet~\cite{Luo2020} & {-} & {-} & - & 6.5 & {-} & 2.28 & 0.734 & 38.1 \\
SISO$_1$~\cite{Wang2020css} & {-} & {-} & - & 5.7 & {-} & 2.40 & 0.748 & 28.7 \\
DNN$_1$+(FCP+DNN$_2$)$\times$2~\cite{Wang2020css} & {-} & {-} & - & 12.7 & 14.1 & 3.25 & 0.899 & 12.8 \\
DNN$_1$+(msFCP+DNN$_2$)$\times$2~\cite{Wang2021FCPwaspaa} & {-} & {-} & - & 13.4 & {-}  & 3.41 & {-} & 10.9 \\
\midrule
Oracle direct-path signal & {-} & {-} & - & $\infty$ & $\infty$ & 4.5 & 1.0 & 6.28 \\
\bottomrule
\end{tabular}
\vspace{-0.2cm}
\end{table}

\subsubsection{Comparison of Loss Functions}\label{loss}

The speaker separation community usually uses SI-SDR as the key evaluation metric and many previous models are trained to optimize SI-SDR.
We also do this in our experiments on WSJ0-2mix in order to compare TF-GridNet with earlier models.
However, using SI-SDR as the loss is known to produce sub-optimal magnitude estimates due to the compensation between estimated magnitude and phase~\cite{Wang2021compensation}, while metrics such as PESQ, eSTOI and WER favor signals with good magnitude estimates.
Based on SMS-WSJ and WHAMR!, in Table~\ref{sms_wsj_results_1ch}, \ref{sms_wsj_results_2ch}, \ref{sms_wsj_results_6ch}, \ref{wharmr_results_1ch} and \ref{wharmr_results_2ch} we make a direct comparison of training TF-GridNet (i.e., DNN$_1$) with the SI-SDR+MC loss in Eq.~(\ref{sinrloss+mc}), Wav+Mag in (\ref{wav+mag}) and Wav+Mag+MC in (\ref{wav+mag+MC}).
We observe that, compared with SI-SDR+MC, Wav+Mag+MC performs better or comparably good in PESQ, eSTOI and WER, and slightly worse in SI-SDR and SDR; and compared with Wav+Mag, it usually performs better.
We will in default use the Wav+Mag+MC loss in the following experiments.

\begin{table}[]
\scriptsize
\centering
\sisetup{table-format=2.2,round-mode=places,round-precision=2,table-number-alignment = center,detect-weight=true,detect-inline-weight=math}
\caption{\textsc{Results on SMS-WSJ (2ch).}}
\vspace{-0.1cm}
\label{sms_wsj_results_2ch}
\setlength{\tabcolsep}{2.5pt}
\begin{tabular}{
l
c %
c %
c
S[table-format=2.1,round-precision=1]
S[table-format=2.1,round-precision=1]
S[table-format=1.2,round-precision=2]
S[table-format=1.3,round-precision=3]
S[table-format=2.2,round-precision=2]
}
\toprule
{\multirow{2}{*}{Systems}} & {\multirow{2}{*}{$\Delta_l$}} & {\multirow{2}{*}{$\Delta_r$}} & {\multirow{2}{*}{Loss}} & {SI-SDR} & {SDR} & {\multirow{2}{*}{PESQ}} & {\multirow{2}{*}{eSTOI}} & {WER} \\
 & & & & {(dB)} & {(dB)} & & & {(\%)} \\
 \midrule
Unprocessed & - & - & - & -5.5 & -0.4 & 1.50 & 0.441 & 78.4 \\
\midrule
DNN$_1$ & {-} & {-} & (\ref{sinrloss+mc}) & 17.8350 & 19.068708 & 3.6723616 & 0.945635 & 8.33 \\
DNN$_1$ & {-} & {-} & (\ref{wav+mag}) & 15.96 & 17.327 & 3.52 & 0.936 & 8.31 \\
DNN$_1$ & {-} & {-} & (\ref{wav+mag+MC}) & 17.7 & 18.9 & 3.68 & 0.950 & 7.68 \\
DNN$_1$+DNN$_2$ & {-} & {-} & (\ref{wav+mag+MC}) & 17.728 & 18.933 & 3.68 & 0.9494756 & 7.90 \\
DNN$_1$+MCMFWF+DNN$_2$ & 0 & 0 & (\ref{wav+mag+MC}) & 17.7790 & 18.9775 & 3.69560 & 0.950037 & 7.84 \\
DNN$_1$+MCMFWF+DNN$_2$ & 29 & 0 & (\ref{wav+mag+MC}) & 19.8989525 & 21.4730 & 3.79483 & 0.9647201 & \bfseries 7.12 \\
DNN$_1$+MCMFWF+DNN$_2$ & 15 & 14 & (\ref{wav+mag+MC}) & \bfseries 20.304 & \bfseries 21.9608 & \bfseries 3.8064 & \bfseries 0.96726 & 7.41 \\
\midrule
DNN$_1$+ConvBF+DNN$_2$ & 29 & {-} & (\ref{wav+mag+MC}) & 19.397328 & 20.861329 & 3.803369 & 0.960734 & 7.52 \\
\midrule
MC-ConvTasNet~\cite{ZhangJisi2020} & {-} & {-} & - & 5.8 & {-} & 2.16 & 0.720 & 45.7 \\
FasNet+TAC~\cite{Luo2020e2e} & {-} & {-} & - & 6.9 & {-} & 2.27 & 0.731 & 34.8  \\
MISO$_1$~\cite{Wang2020css} & {-} & {-} & - & 8.2 & {-} & 2.85 & 0.826 & 17.2 \\
MISO$_1$-BF-MISO$_3$~\cite{Wang2020css} & {-} & {-} & - & 12.7 & {-} & 3.43 & 0.907 &  10.7 \\
\makecell{DNN$_1$+(msFCP$_{\text{MVDR}}$\\+msFCP+DNN$_2$)$\times$2}~\cite{Wang2021FCPwaspaa} & {-} & {-} & - & 15.8 & {-} & 3.71 & {-} & 8.6 \\
\midrule
Oracle direct-path signal & {-} & {-} & - & $\infty$ & $\infty$ & 4.5 & 1.0 & 6.28 \\
\bottomrule
\end{tabular}
\vspace{-0.2cm}
\end{table}

\subsubsection{Comparison in Monaural, Single-DNN Setup}\label{results_reverb_and_noisy_reverb_description}

Table~\ref{sms_wsj_results_1ch} and \ref{wharmr_results_1ch} respectively present the results of TF-GridNet (denoted as DNN$_1$) on the monaural tasks of SMS-WSJ and WHAMR!.
TF-GridNet substantially outperforms competing systems that train a single DNN for separation.
For example, in Table~\ref{sms_wsj_results_1ch} TF-GridNet is 9.2 dB better than DPRNN-TasNet (15.7 vs. 6.5 dB SI-SDR)~\cite{Luo2020} and 10.0 dB better than TCN-DenseUNet based SISO$_1$ (15.7 vs. 5.7 dB SI-SDR)~\cite{Wang2020css}.
To obtain state-of-the-art performance, many previous speaker separation studies tend to use dynamic mixing (DM) to generate more training mixtures.
Their DM results on the monaural task of WHAMR! are listed in the bottom panel of Table~\ref{wharmr_results_1ch}.
Although DM yields slight improvements for previous models, their final performance is still worse than the 10.6 dB SI-SDR result obtained by TF-GridNet without DM (i.e., the DNN$_1$ row).
These results show the effectiveness of TF-GridNet for noisy-reverberant speaker separation.

\begin{table}[]
\scriptsize
\centering
\sisetup{table-format=2.2,round-mode=places,round-precision=2,table-number-alignment = center,detect-weight=true,detect-inline-weight=math}
\caption{\textsc{Results on SMS-WSJ (6ch).}}
\vspace{-0.1cm}
\label{sms_wsj_results_6ch}
\setlength{\tabcolsep}{2.5pt}
\begin{tabular}{
l
c %
c %
c
S[table-format=2.1,round-precision=1]
S[table-format=2.1,round-precision=1]
S[table-format=1.2,round-precision=2]
S[table-format=1.3,round-precision=3]
S[table-format=2.2,round-precision=2]
}
\toprule
{\multirow{2}{*}{Systems}} & {\multirow{2}{*}{$\Delta_l$}} & {\multirow{2}{*}{$\Delta_r$}} & {\multirow{2}{*}{Loss}} & {SI-SDR} & {SDR} & {\multirow{2}{*}{PESQ}} & {\multirow{2}{*}{eSTOI}} & {WER} \\
 & & & & {(dB)} & {(dB)} & & & {(\%)} \\
 \midrule
Unprocessed & {-} & {-} & - & -5.5 & -0.4 & 1.50 & 0.441 & 78.4 \\
\midrule
DNN$_1$ & {-} & {-} & (\ref{sinrloss+mc}) & 19.649174 & 20.993957 & 3.8725276 & 0.9611383543 & 7.63 \\
DNN$_1$ & {-} & {-} & (\ref{wav+mag}) & 19.4 & 20.8 & 3.83 & 0.964 & 6.92 \\
DNN$_1$ & {-} & {-} & (\ref{wav+mag+MC}) & 19.9 & 21.2 & 3.89 & 0.966 & 7.27 \\
DNN$_1$+DNN$_2$ & {-} & {-} & (\ref{wav+mag+MC}) & 19.920 & 21.209 & 3.89 & 0.966 & 7.34 \\
DNN$_1$+MCMFWF+DNN$_2$ & 0 & 0 & (\ref{wav+mag+MC}) & 20.068 & 21.4404 & 3.898 & 0.96655 & 7.28 \\
DNN$_1$+MCMFWF+DNN$_2$ & 9 & 0 & (\ref{wav+mag+MC}) & 22.551951 & 24.578 & 4.03515 & 0.977579 & \bfseries 6.65 \\
DNN$_1$+MCMFWF+DNN$_2$ & 5 & 4 & (\ref{wav+mag+MC}) & \bfseries 22.809367 & \bfseries 24.855 & \bfseries 4.07624 & \bfseries 0.979954 & 6.76 \\
\midrule
DNN$_1$+ConvBF+DNN$_2$ & 9 & {-} & (\ref{wav+mag+MC}) & 21.86570 & 23.6565 & 4.00062 & 0.97459 & 6.74 \\
\midrule
FasNet+TAC~\cite{Luo2020e2e} & {-} & {-} & - & 8.6 & {-} & 2.37 & 0.771 & 29.8 \\
MC-ConvTasNet~\cite{ZhangJisi2020} & {-} & {-} & - & 10.8 & {-} & 2.78 & 0.844 & 23.1 \\
MISO$_1$~\cite{Wang2020css} & {-} & {-} & - & 10.2 & {-} & 3.05 & 0.859 & 14.0 \\
LBT~\cite{Taherian2022LBT} & {-} & {-} & - & 13.2 & 14.8 & 3.33 & 0.910 & 9.6 \\
MISO$_1$-BF-MISO$_3$~\cite{Wang2020css} & {-} & {-} & - & 15.6 & {-} & 3.76 & 0.942 & 8.3 \\
\midrule
Oracle direct-path signal & {-} & {-} & - & $\infty$ & $\infty$ & 4.5 & 1.0 & 6.28 \\
\bottomrule
\end{tabular}
\vspace{-0.2cm}
\end{table}

\subsubsection{Comparison in Multi-Channel, Single-DNN Setup}\label{results_reverb_and_noisy_reverb_description}

Table~\ref{sms_wsj_results_2ch} and \ref{sms_wsj_results_6ch} respectively present the results of TF-GrdiNet based DNN$_1$ for two- and six-channel separation on SMS-WSJ, and Table~\ref{wharmr_results_2ch} reports two-channel results on WHAMR!.
TF-GridNet shows substantially better performance than competing single-DNN approaches.
For example, in Table~\ref{sms_wsj_results_6ch} TF-GridNet obtains 19.9 dB SI-SDR, while FasNet+TAC~\cite{Luo2020e2e}, MC-Conv-TasNet~\cite{ZhangJisi2020}, TCN-DenseUNet~\cite{Wang2020css} and LBT~\cite{Taherian2022LBT} respectively obtain 8.6, 10.8, 10.2 and 13.2 dB.

\subsubsection{Effectiveness of Including MFWF and Post-Filtering}\label{results_MCMFWF_post-filtering}

For the post-filtering network (i.e., DNN$_2$), which is trained in an enhancement way, we use the same configuration as DNN$_1$ but use $B=3$ TF-GridNet blocks.
Although TF-GridNet based DNN$_1$ already exhibits strong separation performance, we observe that using its outputs to compute an MFWF and another TF-GridNet for post-filtering still produces clear improvements.
This can be observed in Table~\ref{sms_wsj_results_2ch} and \ref{sms_wsj_results_6ch} by comparing DNN$_1$+MCMFWF+DNN$_2$, DNN$_1$, and DNN$_1$+DNN$_2$ (which stacks two TF-GridNets but not performing linear filtering in between).
In the monaural case, in Table~\ref{sms_wsj_results_1ch} DNN$_1$+SCMFWF+DNN$_2$ is also better than DNN$_1$.

\begin{table}[]
\scriptsize
\centering
\sisetup{table-format=2.2,round-mode=places,round-precision=2,table-number-alignment = center,detect-weight=true,detect-inline-weight=math}
\caption{\textsc{Results on WHAMR! (1ch).}}
\vspace{-0.1cm}
\label{wharmr_results_1ch}
\setlength{\tabcolsep}{1.7pt}
\begin{tabular}
{l
c %
c %
c
S[table-format=2.1,round-precision=1]
S[table-format=2.1,round-precision=1]
S[table-format=1.2,round-precision=2]
S[table-format=1.3,round-precision=3]
}
\toprule
Systems & $\Delta_l$ & $\Delta_r$ & Loss & {SI-SDR (dB)} & {SDR (dB)} & {PESQ} & {eSTOI} \\
\midrule
Unprocessed & {-} & {-} & - & -6.1 & -3.5 & 1.41 & 0.317 \\
\midrule
DNN$_1$  & {-} & {-}& (\ref{sinrloss+mc}) & 11.0 & 12.1 & 2.69 & 0.790 \\
DNN$_1$  & {-} & {-}& (\ref{wav+mag}) & 10.327445 & 11.39139 & 2.7156155 & 0.786574 \\
DNN$_1$  & {-} & {-}& (\ref{wav+mag+MC}) & 10.640508 & 11.664329 & 2.74717658 & 0.7932153 \\
DNN$_1$+DNN$_2$ & {-} & {-} & (\ref{wav+mag+MC}) & 10.702205 & 11.766915 & 2.7373689 & 0.79382971 \\
DNN$_1$+SCMFWF+DNN$_2$ & 20 & 19 & (\ref{wav+mag+MC}) & \bfseries 11.2242 & \bfseries 12.30882 & \bfseries 2.79482 & \bfseries 0.8077484 \\
\midrule
Conv-TasNet~\cite{Luo2019, Maciejewski2020} & {-} & {-} & -& 2.2 & {-} & {-} & {-} \\
SISO$_1$~\cite{Wang2020css} & {-} & {-} & -& 4.2 & 6.2 & 1.79 & 0.594 \\
3-Stage BLSTM-TasNet~\cite{Maciejewski2020} & {-} & {-} & -& 4.8 & {-} & {-} & {-} \\
Wavesplit~\cite{Zeghidour2020} & {-} & {-} & -& 5.9 & {-} & {-} & {-} \\
Gated DPRNN~\cite{Nachmani2020} & {-} & {-} & -& 6.1 & {-} & {-} & {-} \\
QDPN~\cite{Rixen2022QDPN} & {-} & {-} & -& 7.0 & {-} & {-} & {-} \\
DNN$_1$+(FCP+DNN$_2$)$\times$2~\cite{Wang2021FCPjournal} & {-} & {-} & -& 7.4 & 8.9 & 2.39 & 0.743 \\
\midrule
Wavesplit + DM~\cite{Zeghidour2020} & {-} & {-} & -& 7.1 & 8.7 & {-} & {-} \\
SUDO RM -RF + DM~\cite{Tzinis2020} & {-} & {-} & -& 7.4 & {-} & {-} & {-} \\
SepFormer + DM~\cite{Subakan2021, Subakan2022journal} & {-} & {-} & - & 7.9 & 9.5 & {-} & {-} \\
QDPN + DM~\cite{Rixen2022QDPN} & {-} & {-} & -& 8.3 & {-} & {-} & {-} \\
\bottomrule
\end{tabular}
\vspace{-0.2cm}
\end{table}

\begin{table}[]
\scriptsize
\centering
\sisetup{table-format=2.2,round-mode=places,round-precision=2,table-number-alignment = center,detect-weight=true,detect-inline-weight=math}
\caption{\textsc{Results on WHAMR! (2ch).}}
\vspace{-0.1cm}
\label{wharmr_results_2ch}
\setlength{\tabcolsep}{2pt}
\begin{tabular}{l
c %
c %
c
S[table-format=2.1,round-precision=1]
S[table-format=2.1,round-precision=1]
S[table-format=1.2,round-precision=2]
S[table-format=1.3,round-precision=3]
}
\toprule
Systems & $\Delta_l$ & $\Delta_r$ & Loss & {SI-SDR (dB)} & {SDR (dB)} & {PESQ} & {eSTOI} \\
\midrule
Unprocessed & {-} & {-} & - & -6.1 & -3.5 & 1.41 & 0.317 \\
\midrule
DNN$_1$ & {-} & {-} & (\ref{sinrloss+mc}) & 12.832 & 13.955 & 3.00 & 0.844 \\
DNN$_1$ & {-} & {-} & (\ref{wav+mag}) & 12.0 & 13.2 & 3.01 & 0.839 \\
DNN$_1$ & {-} & {-} & (\ref{wav+mag+MC}) & 12.5 & 13.5 & 3.05 & 0.846 \\
DNN$_1$+DNN$_2$ & {-} & {-} & (\ref{wav+mag+MC}) & 12.5212 & 13.5727 & 3.0542 & 0.84615 \\
DNN$_1$+MCMFWF+DNN$_2$ & 15 & 14 & (\ref{wav+mag+MC}) & \bfseries 13.67305 & \bfseries 14.8236 & \bfseries 3.1639 & \bfseries 0.868275 \\
\midrule
MC-ConvTasNet~\cite{ZhangJisi2020, Zhang2021} & {-} & {-} & - & 6.0 & {-} & {-} & {-} \\
\makecell{MC-ConvTasNet with\\speaker extraction}~\cite{Zhang2021} & {-} & {-} & - & 7.3 & {-} & {-} & {-} \\
\bottomrule
\end{tabular}
\vspace{-0.2cm}
\end{table}

\subsubsection{MFWF vs. Other Linear Filters}\label{comparison_MCMFWF_convbf_wpe}

In Table~\ref{sms_wsj_results_2ch} and \ref{sms_wsj_results_6ch}, we observe that using MCMFWF with both past and future context (i.e., $\Delta_l > 0$ and $\Delta_r > 0$) between DNN$_1$ and DNN$_2$ produces clear improvements over MCMFWF with only past context (i.e., $\Delta_l > 0$ and $\Delta_r = 0$), MCMFWF with no context (i.e., $\Delta_l = 0$ and $\Delta_r = 0$), and convolutional beamformer.
In Table~\ref{sms_wsj_results_1ch}, SCMFWF with both past and future context leads to better scores than WPE as well as SCMFWF with only past context in the single-channel case.

\subsection{Results on WSJ0CAM-DEREVERB}\label{result_dereverb}

\ZQHL{For the rest experiments (including the ones in this subsection and in the next subsection)}, we set $I=4$, $J=2$, and $D=48$. $J$ is increased to $2$ as the sampling rate is 16 kHz.
The other setups are the same as those in the previous subsection. 

\begin{table}[]
\scriptsize
\centering
\sisetup{table-format=2.2,round-mode=places,round-precision=2,table-number-alignment = center,detect-weight=true,detect-inline-weight=math}
\caption{\textsc{Results on WSJ0CAM-DEREVERB (1ch).}}
\vspace{-0.1cm}
\label{dereverb_1ch_results}
\setlength{\tabcolsep}{2.5pt}
\begin{tabular}{l
c %
c %
c
S[table-format=2.1,round-precision=1]
S[table-format=1.2,round-precision=2]
S[table-format=1.3,round-precision=3]
}
\toprule
Systems & $\Delta_l$ & $\Delta_r$ & Loss & {SI-SDR (dB)} & {PESQ} & {eSTOI} \\
\midrule
Unprocessed & {-} & {-} & - & -3.6 & 1.64 & 0.494 \\
\midrule
DNN$_1$  & {-} & {-} & (\ref{wav+mag}) & 16.6 & 3.72 & 0.947 \\
DNN$_1$+DNN$_2$  & {-} & {-}& (\ref{wav+mag}) & 16.9845 & 3.7677 & 0.94836 \\
DNN$_1$+SCMFWF+DNN$_2$  & 20 & 19 & (\ref{wav+mag}) & \bfseries 17.3 & \bfseries 3.78 & \bfseries 0.950 \\
\midrule
SISO$_1$~\cite{Wang2021LowDistortion}  & {-} & {-} & - & 8.4 & 3.12 & 0.868 \\
DNN$_1$+(FCP+DNN$_2$)$\times$2~\cite{Wang2021FCPjournal}  & {-} & {-} & - & 12.7 & 3.46 & {-} \\
\makecell{SISO$_1$+FCP$_{\text{WPE}}$+WPE+SISO$_5$}~\cite{Wang2021LowDistortion}  & {-} & {-} & - & 12.7 & 3.49 & 0.919 \\
\bottomrule
\end{tabular}
\vspace{-0.2cm}
\end{table}

\begin{table}[]
\scriptsize
\centering
\sisetup{table-format=2.2,round-mode=places,round-precision=2,table-number-alignment = center,detect-weight=true,detect-inline-weight=math}
\caption{\textsc{Results on WSJ0CAM-DEREVERB (8ch).}}
\vspace{-0.1cm}
\label{dereverb_8ch_results}
\setlength{\tabcolsep}{2.5pt}
\begin{tabular}{l
c %
c %
c
S[table-format=2.1,round-precision=1]
S[table-format=1.2,round-precision=2]
S[table-format=1.3,round-precision=3]
}
\toprule
Systems & $\Delta_l$ & $\Delta_r$ & Loss & {SI-SDR (dB)} & {PESQ} & {eSTOI} \\
\midrule
Unprocessed & {-} & {-} & - & -3.6 & 1.64 & 0.494 \\
\midrule
DNN$_1$ & {-} & {-} & (\ref{wav+mag}) & 19.9 & 3.95 & 0.971 \\
DNN$_1$+DNN$_2$ & {-} & {-} & (\ref{wav+mag}) & 20.288 & 4.00 & 0.972 \\
DNN$_1$+MCMFWF+DNN$_2$ & 4 & 3 & (\ref{wav+mag}) & \bfseries 21.2 & \bfseries 4.02 & \bfseries 0.975 \\
\midrule
MISO$_1$~\cite{Wang2021LowDistortion} & {-} & {-} & - & 11.3 & 3.49 & 0.921 \\
\makecell[l]{MISO$_1$+FCP$_{\text{mWMPDR}_{\text{WPE}}}$+\\\,\,\,\,mWMPDR$_{\text{WPE}}$+WPE+MISO$_{10}$}~\cite{Wang2021LowDistortion} & {-} & {-} & - & 18.2 & 3.98 & 0.967 \\
\bottomrule
\end{tabular}
\vspace{-0.2cm}
\end{table}

Table~\ref{dereverb_1ch_results} and \ref{dereverb_8ch_results} respectively present the results of using TF-GridNet for one- and eight-channel dereverberation.
Trained to perform complex spectral mapping, DNN$_1$ based on TF-GridNet achieves substantially better performance than SISO$_1$ (16.6 vs. 8.4 dB SI-SDR in Table~\ref{dereverb_1ch_results}) and MISO$_1$ (19.9 vs. 11.3 dB SI-SDR in Table~\ref{dereverb_8ch_results}) proposed in~\cite{Wang2021LowDistortion}, which also uses complex spectral mapping but with TCN-DenseNet.
With beamforming and post-filtering, DNN$_1$+MCMFWF+DNN$_2$ based on TF-GridNet also shows better results than the competing approach (21.2 vs. 18.2 dB SI-SDR) in~\cite{Wang2021LowDistortion}, which uses two TCN-DenseUNets with a composition of linear filters.

From the last two rows of Table~\ref{dereverb_8ch_results}, we notice that, based on TCN-DenseUNet, using complicated sub-band linear filtering followed by post-filtering (i.e., the last row) produces large improvement over MISO$_1$ (18.2 vs. 11.3 dB SI-SDR)~\cite{Wang2021LowDistortion}.
This indicates that the sub-band linear filters can model what TCN-DenseUNet, which performs full-band modeling, is not good at modeling.
In comparison, using a single TF-GridNet alone is already better than the last two rows (i.e., 19.9 vs. 11.3 and 18.2 dB SI-SDR) and the improvement brought by beamforming and post-filtering is not large (21.2 vs. 19.9 dB SI-SDR).
This indicates that TF-GridNet could, to a large extent, model what sub-band linear filters complement to full-band models, likely through the sub-band temporal modules.

\subsection{Results on L3DAS22}\label{result_L3DAS_description}

L3DAS22 requires participants to estimate the dry source signal.
Following Eq.~(\ref{wav+mag}), we define the loss as
\begin{align}\label{wav+mag+GEQ}
&\mathcal{L}_{\text{Wav+Mag,GEQ}} =
\sum\nolimits_{c=1}^C \Big(\frac{1}{N} \| \hat{\alpha}^{(c)} \hat{o}(c) - o(c) \|_1 + \nonumber \\ 
&\,\,\,\,\,\,\,\,\,\,\,\,\,\,\,\,\frac{1}{T\times F} \Big\| \Big|\text{STFT}(\hat{\alpha}^{(c)}\hat{o}(c))\Big| - \Big|\text{STFT}(o(c))\Big| \Big\|_1\Big),
\end{align}
where $o(c)$ denotes the dry source signal of speaker $c$ (see our physical model in Eq.~(\ref{eq:phymodel_time})) and $\hat{\alpha}^{(c)} = {{\text{argmin}}}_{\alpha}\,\| \alpha \hat{o}^{(c)} - o^{(c)} \|_2^2=(\hat{o}^{(c)})^{\T}o^{(c)}/(\hat{o}^{(c)})^{\T}\hat{o}^{(c)}$ is a gain equalization (GEQ) factor~\cite{LeRoux2019, Lu2022} that allows estimated speech to have an energy level different from target speech.

Table~\ref{l3das_results} reports the results.
A single TF-GridNet (i.e., DNN$_1$) already outperforms our winning solution~\cite{Lu2022} and the rest 16 submissions (see this link\footnote{See \url{https://www.l3das.com/icassp2022/results.html}} for the challenge ranking), including the runner-up system~\cite{Zhang2022BaiduSpeechL3DAS}, whose monaural version~\cite{Zhang2022Axial} won the recent DNS2022 and AEC2022 challenges.

Including beamforming and post-filtering yields further improvement.
Here MCMFWF is computed in a way similarly to Eq.~(\ref{MCMFWF}), but we project the far-field B-format Ambisonic mixture to the dry source signal estimated by DNN$_1$ so that the beamforming result can be potentially time-aligned with the dry target, if the dry target estimated by DNN$_1$ is reasonably good, which is the case from the DNN$_1$ row.
In comparison, DNN-supported convolutional beamformer cannot produce an estimate time-aligned with the dry source, and how to modify it to deal with B-format Ambisonic signals is unknown.

\begin{table}[]%
\scriptsize
\centering
\sisetup{table-format=2.2,round-mode=places,round-precision=2,table-number-alignment = center,detect-weight=true,detect-inline-weight=math}
\caption{\textsc{Results on L3DAS22 3D Speech Enhancement Task (8ch).}}
\vspace{-0.1cm}
\label{l3das_results}
\setlength{\tabcolsep}{1.5pt}
\begin{tabular}{l
c %
S[table-format=1.1,round-precision=1]
c %
S[table-format=2.2,round-precision=2]
S[table-format=1.3,round-precision=3]
S[table-format=1.3,round-precision=3]
}
\toprule
{\multirow{2}{*}{Systems}} & {\multirow{2}{*}{$\Delta_l/\Delta_r$}} & {\#params} & {\multirow{2}{*}{Loss}} & {\multirow{2}{*}{WER (\%)}} & {\multirow{2}{*}{STOI}} & {\multirow{2}{*}{Task1Metric}} \\
 & & {(M)} & & & & \\
\midrule
DNN$_1$ & {-} & 5.6 & (\ref{wav+mag+GEQ}) & 1.6776608371442375 & 0.98774326113973 & 0.9854833263841438 \\
DNN$_1$+DNN$_2$ & {-} & 9.8 & (\ref{wav+mag+GEQ}) & 1.625296164170487 & 0.9888854462821834 & 0.9863162423202393 \\
DNN$_1$+MCMFWF+DNN$_2$ & {$4/3$} & 9.8 & (\ref{wav+mag+GEQ}) & \bfseries 1.2931165717770235 & \bfseries 0.9937596966842833 & \bfseries 0.9904142654832564 \\
\midrule
Winner (ESP-SE)~\cite{Lu2022} & {-} & 13.8 & - & 1.89 & 0.987 & 0.984 \\
Runner-up (BaiduSpeech)~\cite{Zhang2022BaiduSpeechL3DAS} & {-} & {-} & - & 2.50 & 0.975 & 0.975 \\
3nd-place (PCG-AIID)~\cite{Li2022PCGL3DAS} & {-} & {-} & - & 3.20 & 0.972 & 0.970 \\
Challenge baseline~\cite{ren2021neural} & {-} & 5.5 & - & 21.2 & 0.878 & 0.833 \\
\bottomrule
\end{tabular}
\vspace{-0.2cm}
\end{table}

\begin{table*}[t]
\scriptsize
\centering
\sisetup{table-format=2.2,round-mode=places,round-precision=2,table-number-alignment =center,detect-weight=true,detect-inline-weight=math}
\caption{\textsc{Ablation Results of Computation Cost and Separation Performance on WSJ0-2mix.}}
\vspace{-0.1cm}
\label{result_computation_vs_performance}
\setlength{\tabcolsep}{1.75pt}
\begin{tabular}{
r %
c %
c %
S[table-format=1,round-precision=0] %
S[table-format=2.1,round-precision=1] %
c %
S[table-format=2.1,round-precision=1] %
c %
S[table-format=3.1,round-precision=1] %
S[table-format=4.1,round-precision=1] %
S[table-format=4.1,round-precision=1] %
S[table-format=3.1,round-precision=1] %
S[table-format=3.1,round-precision=1] %
S[table-format=3.1,round-precision=1] %
S[table-format=5.1,round-precision=1] %
}
\toprule

 & & &  & & & & & & \multicolumn{2}{c}{Memory cost (GPU)} & \multicolumn{3}{c}{Speed (GPU)} & \multicolumn{1}{c}{Speed (CPU)} \\
 \cmidrule(lr{9pt}){10-11}
 \cmidrule(lr){12-14}
  \cmidrule(lr){15-15}
{\multirow{2}{*}{\rotatebox[origin=c]{0}{Row}}} & {\multirow{2}{*}{Systems}} & {Unfold+LN or} & {\multirow{2}{*}{$L/D/I/J/H$}} & {\#params} & {\multirow{2}{*}{Loss}} & {SI-SDRi} & {Window/hop} & {\multirow{2}{*}{GMAC/s}} & {Forward} & {Backward} & {Training} & {Forward} & {Backward} & {Forward}\\
 & & {LN+Unfold} & & {(M)} & & {(dB)} & sizes (ms) & & {(MB/seg)} & {(MB/seg)} & {(min/epoch)} & {(ms/seg)} & {(ms/seg)}  & {(ms/seg)} \\

\midrule

1 & TF-GridNet & {LN+Unfold} & {$4/64/4/1/256$} & 14.5 & (\ref{sinrloss+mc}) & 23.50169992613451 & {$32/8$} & 231.143710181 & 4918.760448 & 13663.19104 & 151.4 & 450.7961632653062 & 832.9273469387756 & 18440.646734693877 \\
2 & TF-GridNet & {LN+Unfold} & {$4/48/4/1/192$} & 8.2 & (\ref{sinrloss+mc}) & 23.175252076899632 & {$32/8$} & 131.107429049 & 4581.890048 & 10871.365120000002 & 119.51666666666667 & 1438.9504693877552 & 609.6977959183674 & 12169.194510204079 \\
3 & TF-GridNet & {LN+Unfold} & {$4/48/4/1/192$} & 8.2 & (\ref{sinrloss+mc}) & 23.160769594814628  & {$16/8$} & 65.993780711 & 2520.128 & 5803.284991999999 & 71.60833333333333 & 756.4623469387757 & 403.9848163265306 & 6383.157673469388 \\
4 & TF-GridNet & {LN+Unfold} & {$4/96/2/2/192$} & 8.4 & (\ref{sinrloss+mc}) & 22.242037795156552 & {$16/8$} & 36.209552915 & 2282.6091520000004 & 3853.7318399999995 & 24.150000000000002 & 363.9553265306123 & 188.81669387755102 & 3563.9715306122444 \\
5 & TF-GridNet & {LN+Unfold} & {$4/64/3/3/192$} & 8.2 & (\ref{sinrloss+mc}) & 21.25016786118775 & {$16/8$} & 24.421761515 & 1620.5619199999996 & 2686.8902347755097 & 18.029166666666665 & 247.18967346938774 & 132.0634693877551 & 2428.9181020408164 \\
6 & TF-GridNet & {LN+Unfold} & {$4/48/4/4/192$} & 8.2 & (\ref{sinrloss+mc}) & 20.62936843696516 & {$16/8$} & 19.243329527 & 1318.713856 & 2189.3918719999997 & 15.341666666666667 & 192.99636734693877 & 106.40359183673469 & 1930.8377551020408 \\
7 & TF-GridNet & {LN+Unfold} & {$4/32/4/4/128$} & 3.7 & (\ref{sinrloss+mc}) & 19.95624715641917 & {$16/8$} & 9.516136835 & 956.2675200000001 & 1590.818816 & 12.61021505376344 & 27.77730612244898 & 58.04616326530613 & 1206.7380816326531 \\
8 & TF-GridNet & {LN+Unfold} & {$4/24/4/4/\,\,\,96$} & 2.1 & (\ref{sinrloss+mc}) & 18.89557991660821 & {$16/8$} & 6.092626313 & 785.934848 & 1299.9731199999999 & 11.363333333333333 & 17.029020408163266 & 62.06355102040816 & 903.7123061224489 \\

\hdashline

9 & TF-GridNet & {LN+Unfold} & {$4/88/2/2/172$} & 6.8 & (\ref{sinrloss+mc}) & 22.023998906963193 & {$16/8$} & 29.825623145 & 2093.17632 & 3495.988736 & 23.06140350877193 & 325.50710204081633 & 192.60008163265303 & 3189.9648367346945 \\

\midrule

10 & Conv-TasNet \cite{Luo2019} & {-} & {-} & 5.1 & {-} & 15.6 & {$2/1$} & 5.085 & 1326.802432 & 1445.221888 & 7.4229166666666675 & 88.09177551020409 & 26.021183673469388 & 602.1929387755102 \\
11 & DPRNN \cite{Luo2020} & {-} & {-} & 2.6 & {-} & 18.8 & {$\,\,0.25/0.125$} & 42.2025 & 2927.4316799999997 & 7049.58208 & 28.64333333333333 & 121.7971224489796 & 228.56626530612246 & 4109.355387755102 \\
12 & SepFormer \cite{Subakan2021} & {-} & {-} & 26.0 & {-} & 20.4 & {$2/1$} & 59.4875 & 5465.734144 & 5703.890944 & {\,\,\,\,N/A} & 82.24769387755103 & 135.004693877551 & 5734.073816326531 \\
13 & TFPSNet \cite{Yang2022TFPSNet} & {-} & {-} & 2.7 & {-} & 21.1 & {$32/16$} & 29.649346599 & 4869.416448 & 7182.0088319999995 & 35.162962962962965 & 77.35273469387755 & 184.54502040816323 & 4486.812897959184 \\

\bottomrule

\multicolumn{15}{l}{\textit{Notes}:}\\
\multicolumn{15}{l}{(a) TF-GridNet is configured to always include the self-attention module.}\\
\multicolumn{15}{l}{(b) GMAC/s is computed based on a 4-second segment and batch size 1.}\\
\multicolumn{15}{l}{(c) For the memory cost on GPU in the forward/backward pass, we report it in megabytes (MB) per 4-second segment.}\\
\multicolumn{15}{l}{(d) For the training speed, we report the time in minutes to finish an epoch with 20,000 4-second segments on an NVIDIA A100 GPU with $40$ GB memory. The batch size is}\\
\multicolumn{15}{l}{\quad\,\,\,set so that nearly all the GPU memory is utilized. The result of SepFormer is not available, because, in each training step, it is designed to model an entire mixture.}\\
\multicolumn{15}{l}{(e) For the forward (and backward) speed on GPU, we report the averaged time in millisecond (ms) taken to finish processing 4-second segments with a batch size of $1$. An}\\
\multicolumn{15}{l}{\quad\,\,\,NVIDIA Tesla V100 GPU with $32$ GB memory is used.}\\
\multicolumn{15}{l}{(f) For the forward speed on CPU, we also report the averaged time in millisecond to process 4-second segments with a batch of $1$. A single core of a CPU, Intel(R) Xeon(R)}\\
\multicolumn{15}{l}{\quad\,\,\,Gold 6242 CPU @ 2.80GHz, is used.}\\
\multicolumn{15}{l}{(g) For the columns on forward/backward memory and speed, we use Pytorch v2.0.1 and \textit{torch.profiler} for profiling.}\\
\end{tabular}
\vspace{-0.2cm}
\end{table*}

\ZQHL{
\subsection{Computation Cost vs. Separation Performance}\label{cost_perf}

Although this paper focuses on the separation performance rather than computation cost, this subsection varies the computation cost of TF-GridNet and reports the separation performance on WSJ0-2mix in Table \ref{result_computation_vs_performance} (see the notes below the table for how we calculate the computation cost).
The computation cost can be controlled by increasing the stride size $J$ (so that the sequence length is reduced), reducing the overlap between consecutive frames, and reducing the hidden dimensions of BLSTMs, $H$, as well as the embedding dimension $D$.

In row 1, the model obtaining the $23.5$ dB result (i.e., the best result in Table \ref{comparison_with_others}) is very costly, requiring $231.1$ GMAC/s which is quite high.
In row 2, we reduce the hidden units in BLSTMs, $H$, from $256$ to $196$ and the embedding dimension $D$ from $64$ to $48$. 
This reduces GMAC/s to $131.1$, while the performance degrades slightly to $23.2$ dB. 
In row 3, we decrease the window size from $32$ to $16$ ms, reducing GMAC/s by almost half to $66.0$, as the number of frequencies is cut by around half.
The performance, nonetheless, remains at $23.2$ dB.
In row 4-6, we increase the stride size $J$ from $1$ to $2$, $3$ and $4$ respectively, set the kernel size $I$ equal to $J$, and set $D$ such that $D\times I=H$.
These reduce GMAC/s from $66.0$ gradually down to $19.2$, as the sequence length the BLSTMs need to model becomes shorter.
In row 7, we reduce the hidden units in BLSTMs, $H$, from $192$ to $128$. The computation is further reduced to $9.5$ GMAC/s, and the performance is at $20.0$ dB.
In row 8, we reduce the hidden units in BLSTMs, $H$, from $128$ to $96$, resulting in $6.1$ GMAC/s. The model can still obtain $18.9$ dB.

Table \ref{result_computation_vs_performance} provides the training speed on a modern GPU in terms of the number of minutes taken to finish an epoch.
Although the models in the first two rows take a long time (i.e., $119.5$ and $231.1$ minutes) to complete an epoch, all the other configurations have a reasonable training time per epoch.

Table \ref{result_computation_vs_performance} compares the computation cost of several other representative models with that of TF-GridNet.
TF-GridNet obtains competitive performance, given limited computation cost.
For example, TF-GridNet in row 9 obtains better separation performance than TFPSNet \cite{Yang2022TFPSNet} in row 13 (i.e, $22.0$ vs. $21.1$ dB SI-SDRi), using similar GMAC/s (i.e., $29.8$ vs. $29.6$), less memory, and exhibiting faster inference speed on CPU.
Compared with a representative time-domain model, DPRNN \cite{Luo2020}, shown in row 11, the TF-GridNets from row 4 to 9 can all obtain better separation performance using fewer GMAC/s. 

These results suggest that TF-GridNet can be configured, in a flexible way, to use a reasonable amount of computation and achieve a reasonable separation performance.
}

\section{Conclusion}\label{conclusion}

We have proposed TF-GridNet, a
DNN architecture modeling complex T-F spectrograms, for single- and multi-channel speech separation.
By integrating full- and sub-band modeling inside TF-GridNet and outside through beamforming and post-filtering, the proposed systems achieve state-of-the-art performance for speech separation in noisy-reverberant conditions on multiple public datasets.
\ZQHL{Our future research will extend TF-GridNet for real-time, online speech separation, building upon our preliminary investigations \cite{Cornell2022Clarity, cornell2023multi, Wang2023ICASSPLowLat} which have shown promising results.}

TF-GridNet obtains a state-of-the-art $23.5$ dB SI-SDRi on WSJ0-2mix, \ZQHL{and it can be configured to use a reasonable amount of computation and achieve a reasonable separation performance.}
These resulte highlights the strong performance of T-F domain models also for anechoic speaker separation, suggesting that T-F domain methods modeling complex representations, which implicitly perform phase estimation by predicting target RI components simultaneously, are not sub-optimal compared to time-domain approaches for the task of anechoic speaker separation.
The performance differences between these two approaches observed in earlier studies could mainly result from their differences in DNN architectures. %

In closing, we emphasize that (i) the patterns of speech spectrograms vary with frequency \ZQHL{but, within each sub-band, some patterns such as spatial and reverberation patterns are relatively stable along time; and (ii) full-band or sub-band modeling alone is likely not capable of sufficiently modeling such patterns.}
Our proposed ways to integrate them exhibit excellent performance in our experiments.
The meta-idea of integrated full- and sub-band modeling, we believe, would motivate the design of many new algorithms in future research on neural speech separation.

\section{Acknowledgments}\label{ack}

We would like to thank Dr. Wangyou Zhang at SJTU for generously sharing his reproduced code of TFPSNet.
This research is part of the Delta research computing project, which is supported by the National Science Foundation (Award OCI $2005572$) and the State of Illinois.
Delta is a joint effort of the University of Illinois at Urbana-Champaign and its National Center for Supercomputing Applications.
We also gratefully acknowledge the support of NVIDIA Corporation with the donation of the RTX 8000 GPUs used in this research.

\bibliographystyle{IEEEtran}
\bibliography{references.bib}

% Generated by IEEEtran.bst, version: 1.14 (2015/08/26)
\begin{thebibliography}{10}
\providecommand{\url}[1]{#1}
\csname url@samestyle\endcsname
\providecommand{\newblock}{\relax}
\providecommand{\bibinfo}[2]{#2}
\providecommand{\BIBentrySTDinterwordspacing}{\spaceskip=0pt\relax}
\providecommand{\BIBentryALTinterwordstretchfactor}{4}
\providecommand{\BIBentryALTinterwordspacing}{\spaceskip=\fontdimen2\font plus
\BIBentryALTinterwordstretchfactor\fontdimen3\font minus
  \fontdimen4\font\relax}
\providecommand{\BIBforeignlanguage}[2]{{%
\expandafter\ifx\csname l@#1\endcsname\relax
\typeout{** WARNING: IEEEtran.bst: No hyphenation pattern has been}%
\typeout{** loaded for the language `#1'. Using the pattern for}%
\typeout{** the default language instead.}%
\else
\language=\csname l@#1\endcsname
\fi
#2}}
\providecommand{\BIBdecl}{\relax}
\BIBdecl

\bibitem{WDLreview}
D.~Wang and J.~Chen, ``{Supervised Speech Separation Based on Deep Learning: An
  Overview},'' \emph{IEEE/ACM Trans. Audio, Speech, Lang. Process.}, vol.~26,
  no.~10, pp. 1702--1726, 2018.

\bibitem{Hershey2016}
J.~R. Hershey, Z.~Chen, J.~{Le Roux}, and S.~Watanabe, ``{Deep Clustering:
  Discriminative Embeddings for Segmentation and Separation},'' in \emph{Proc.
  ICASSP}, 2016, pp. 31--35.

\bibitem{Kolbak2017}
M.~Kolb{\ae}k, D.~Yu, Z.-H. Tan, and J.~Jensen, ``{Multitalker Speech
  Separation with Utterance-Level Permutation Invariant Training of Deep
  Recurrent Neural Networks},'' \emph{IEEE/ACM Trans. Audio, Speech, Lang.
  Process.}, vol.~25, no.~10, pp. 1901--1913, 2017.

\bibitem{Isik2016}
Y.~Isik, J.~{Le Roux} \emph{et~al.}, ``{Single-Channel Multi-Speaker Separation
  using Deep Clustering},'' in \emph{Proc. Interspeech}, 2016, pp. 545--549.

\bibitem{Wang2018AlternativeObejectives}
Z.-Q. Wang, J.~{Le Roux}, and J.~R. Hershey, ``{Alternative Objective Functions
  for Deep Clustering},'' in \emph{Proc. ICASSP}, 2018, pp. 686--690.

\bibitem{WZQe2eMISI2018}
Z.-Q. Wang, J.~{Le Roux}, D.~Wang, and J.~R. Hershey, ``{End-to-End Speech
  Separation with Unfolded Iterative Phase Reconstruction},'' in \emph{Proc.
  Interspeech}, 2018, pp. 2708--2712.

\bibitem{Wang2019Trigonometric}
Z.-Q. Wang, K.~Tan, and D.~Wang, ``{Deep Learning Based Phase Reconstruction
  for Speaker Separation: A Trigonometric Perspective},'' in \emph{Proc.
  ICASSP}, 2019, pp. 71--75.

\bibitem{Liu2019DeepCASA}
Y.~Liu and D.~Wang, ``{Divide and Conquer: A Deep CASA Approach to
  Talker-Independent Monaural Speaker Separation},'' \emph{IEEE/ACM Trans.
  Audio, Speech, Lang. Process.}, vol.~27, no.~12, pp. 2092--2102, 2019.

\bibitem{Luo2017TasNet}
Y.~Luo and N.~Mesgarani, ``{TasNet: Time-Domain Audio Separation Network for
  Real-Time, Single-Channel Speech Separation},'' in \emph{Proc. ICASSP}, nov
  2017, pp. 697--700.

\bibitem{Luo2018TasNetRealTime}
------, ``{Real-Time Single-Channel Dereverberation and Separation with
  Time-Domain Audio Separation Network},'' in \emph{Proc. Interspeech}, 2018,
  pp. 342--346.

\bibitem{Luo2019}
------, ``{Conv-TasNet: Surpassing Ideal Time-Frequency Magnitude Masking for
  Speech Separation},'' \emph{IEEE/ACM Trans. Audio, Speech, Lang. Process.},
  vol.~27, no.~8, pp. 1256--1266, 2019.

\bibitem{Lam2020MBT}
M.~W. Lam, J.~Wang, D.~Su, and D.~Yu, ``{Mixup-Breakdown: A Consistency
  Training Method for Improving Generalization of Speech Separation Models},''
  in \emph{Proc. ICASSP}, 2020, pp. 6374--6378.

\bibitem{Shi2019FurcaNeXt}
L.~Zhang \emph{et~al.}, ``{FurcaNeXt: End-to-End Monaural Speech Separation
  with Dynamic Gated Dilated Temporal Convolutional Networks},'' in \emph{Proc.
  ICMM}, 2020, pp. 653--665.

\bibitem{Tzinis2020}
E.~Tzinis, Z.~Wang, and P.~Smaragdis, ``{Sudo RM -RF: Efficient Networks for
  Universal Audio Source Separation},'' in \emph{Proc. MLSP}, 2020.

\bibitem{Luo2020}
Y.~Luo, Z.~Chen, and T.~Yoshioka, ``{Dual-Path RNN: Efficient Long Sequence
  Modeling for Time-Domain Single-Channel Speech Separation},'' in \emph{Proc.
  ICASSP}, 2020, pp. 46--50.

\bibitem{Nachmani2020}
E.~Nachmani \emph{et~al.}, ``{Voice Separation with An Unknown Number of
  Multiple Speakers},'' in \emph{Proceedings of ICML}, 2020, pp. 7121--7132.

\bibitem{Chen2020DPTnet}
J.~Chen, Q.~Mao, and D.~Liu, ``{Dual-Path Transformer Network: Direct
  Context-Aware Modeling for End-to-End Monaural Speech Separation},'' in
  \emph{Proc. Interspeech}, 2020, pp. 2642--2646.

\bibitem{Zhu2021}
Y.~Zhu, X.~Zheng, X.~Wu, W.~Liu, L.~Pi, and M.~Chen, ``{DPTCN-ATPP: Multi-Scale
  End-To-end Modeling for Single-Channel Speech Separation},'' in \emph{Proc.
  ICCIS}, 2021, pp. 39--44.

\bibitem{Subakan2021}
C.~Subakan, M.~Ravanelli, S.~Cornell \emph{et~al.}, ``{Attention Is All You
  Need In Speech Separation},'' in \emph{Proc. ICASSP}, 2021, pp. 21--25.

\bibitem{Lam2021}
M.~W.~Y. Lam, J.~Wang, D.~Su, and D.~Yu, ``{Sandglasset: A Light
  Multi-Granularity Self-Attentive Network for Time-Domain Speech
  Separation},'' in \emph{Proc. ICASSP}, 2021, pp. 5759--5763.

\bibitem{Zeghidour2020}
N.~Zeghidour and D.~Grangier, ``{Wavesplit: End-to-End Speech Separation by
  Speaker Clustering},'' \emph{IEEE/ACM Trans. Audio, Speech, Lang. Process.},
  vol.~29, pp. 2840--2849, 2021.

\bibitem{Qian2022}
S.~Qian, L.~Gao, H.~Jia, and Q.~Mao, ``{Efficient Monaural Speech Separation
  With Multiscale Time-Delay Sampling},'' in \emph{Proc. ICASSP}, 2022, pp.
  6847--6851.

\bibitem{Rixen2022}
J.~Rixen and M.~Renz, ``{SFSRNet: Super-Resolution for Single-Channel Audio
  Source Separation},'' in \emph{Proceedings of AAAI}, 2022.

\bibitem{Rixen2022QDPN}
------, ``{QDPN - Quasi-Dual-Path Network for Single-Channel Speech
  Separation},'' in \emph{Proc. Interspeech}, 2022, pp. 5353--5357.

\bibitem{Yang2022TFPSNet}
L.~Yang, W.~Liu, and W.~Wang, ``{TFPSNet: Time-Frequency Domain Path Scanning
  Network for Speech Separation},'' in \emph{Proc. ICASSP}, 2022, pp.
  6842--6846.

\bibitem{Le2021DPCRN}
X.~Le, H.~Chen, K.~Chen, and J.~Lu, ``{DPCRN: Dual-Path Convolution Recurrent
  Network for Single Channel Speech Enhancement},'' in \emph{Proc.
  Interspeech}, vol.~2, 2021, pp. 821--825.

\bibitem{Dang2022DPTFSNet}
F.~Dang, H.~Chen, and P.~Zhang, ``{DPT-FSNet:Dual-Path Transformer Based
  Full-Band and Sub-Band Fusion Network for Speech Enhancement},'' in
  \emph{Proc. ICASSP}, 2022, pp. 6857--6861.

\bibitem{Wang2022GridNet}
Z.-Q. Wang, S.~Cornell, S.~Choi \emph{et~al.}, ``{TF-GridNet: Making
  Time-Frequency Domain Models Great Again for Monaural Speaker Separation},''
  in \emph{arXiv preprint arXiv:2209.03952}, 2022.

\bibitem{Williamson2016}
D.~S. Williamson, Y.~Wang, and D.~Wang, ``{Complex Ratio Masking for Monaural
  Speech Separation},'' \emph{IEEE/ACM Trans. Audio, Speech, Lang. Process.},
  pp. 483--492, 2016.

\bibitem{Fu2017}
S.-W. Fu, T.-Y. Hu, Y.~Tsao, and X.~Lu, ``{Complex Spectrogram Enhancement by
  Convolutional Neural Network with Multi-Metrics Learning},'' in \emph{Proc.
  MLSP}, 2017, pp. 1--6.

\bibitem{Tan2020}
K.~Tan and D.~Wang, ``{Learning Complex Spectral Mapping With Gated
  Convolutional Recurrent Networks for Monaural Speech Enhancement},''
  \emph{IEEE/ACM Trans. Audio, Speech, Lang. Process.}, vol.~28, pp. 380--390,
  2020.

\bibitem{Wang2020CSMDereverbJournal}
Z.-Q. Wang and D.~Wang, ``{Deep Learning Based Target Cancellation for Speech
  Dereverberation},'' \emph{IEEE/ACM Trans. Audio, Speech, Lang. Process.},
  vol.~28, pp. 941--950, 2020.

\bibitem{Wang2020chime}
Z.-Q. Wang, P.~Wang, and D.~Wang, ``{Complex Spectral Mapping for Single- and
  Multi-Channel Speech Enhancement and Robust ASR},'' \emph{IEEE/ACM Trans.
  Audio, Speech, Lang. Process.}, vol.~28, pp. 1778--1787, 2020.

\bibitem{Wang2020dMCCSMconference}
Z.-Q. Wang and D.~Wang, ``{Multi-Microphone Complex Spectral Mapping for Speech
  Dereverberation},'' in \emph{Proc. ICASSP}, 2020, pp. 486--490.

\bibitem{Wang2020css}
Z.-Q. Wang \emph{et~al.}, ``{Multi-Microphone Complex Spectral Mapping for
  Utterance-Wise and Continuous Speech Separation},'' \emph{IEEE/ACM Trans.
  Audio, Speech, Lang. Process.}, vol.~29, pp. 2001--2014, 2021.

\bibitem{Wang2021LowDistortion}
Z.-Q. Wang, G.~Wichern, and J.~{Le Roux}, ``{Leveraging Low-Distortion Target
  Estimates for Improved Speech Enhancement},'' \emph{arXiv preprint
  arXiv:2110.00570}, 2021.

\bibitem{Wang2021FCPjournal}
Z.-Q. Wang \emph{et~al.}, ``{Convolutive Prediction for Monaural Speech
  Dereverberation and Noisy-Reverberant Speaker Separation},'' \emph{IEEE/ACM
  Trans. Audio, Speech, Lang. Process.}, vol.~29, pp. 3476--3490, 2021.

\bibitem{Tan2022NSF}
K.~Tan, Z.-Q. Wang \emph{et~al.}, ``{Neural Spectrospatial Filtering},''
  \emph{IEEE/ACM Trans. Audio, Speech, Lang. Process.}, vol.~30, pp. 605--621,
  2022.

\bibitem{Wang2021FCPwaspaa}
Z.-Q. Wang, G.~Wichern, and J.~{Le Roux}, ``{Convolutive Prediction for
  Reverberant Speech Separation},'' in \emph{Proc. WASPAA}, 2021, pp. 56--60.

\bibitem{LeRoux2019}
J.~{Le Roux}, S.~Wisdom, H.~Erdogan, and J.~R. {Hershey}, ``{SDR - Half-Baked
  or Well Done?}'' in \emph{Proc. ICASSP}, 2019, pp. 626--630.

\bibitem{Drude2019}
L.~Drude, J.~Heitkaemper \emph{et~al.}, ``{SMS-WSJ: Database, Performance
  Measures, and Baseline Recipe for Multi-Channel Source Separation and
  Recognition},'' in \emph{arXiv preprint arXiv:1910.13934}, 2019.

\bibitem{Maciejewski2020}
M.~MacIejewski \emph{et~al.}, ``{WHAMR!: Noisy and Reverberant Single-Channel
  Speech Separation},'' in \emph{Proc. ICASSP}, 2020, pp. 696--700.

\bibitem{Lu2022}
Y.-J. Lu, S.~Cornell, X.~Chang, W.~Zhang, C.~Li, Z.~Ni, Z.-Q. Wang, and
  S.~Watanabe, ``{Towards Low-Distortion Multi-Channel Speech Enhancement: The
  ESPNet-SE Submission to The L3DAS22 Challenge},'' in \emph{Proc. ICASSP},
  2022, pp. 9201--9205.

\bibitem{Guizzo2022L3DAS}
E.~Guizzo, C.~Marinoni, M.~Pennese \emph{et~al.}, ``{L3DAS22 Challenge:
  Learning 3D Audio Sources in a Real Office Environment},'' in \emph{Proc.
  ICASSP}, 2022, pp. 9186--9190.

\bibitem{Lu2022ESPNetSE++}
Y.-J. Lu, X.~Chang, C.~Li, W.~Zhang \emph{et~al.}, ``{ESPnet-SE++: Speech
  Enhancement for Robust Speech Recognition,Translation, and Understanding},''
  in \emph{Proc. Interspeech}, 2022.

\bibitem{Paszke2019}
A.~Paszke, S.~Gross, F.~Massa \emph{et~al.}, ``{PyTorch: An Imperative Style,
  High-Performance Deep Learning Library},'' in \emph{Proc. NIPS}, vol.~32,
  2019.

\bibitem{Wang2018combiningspectralspatial}
Z.-Q. Wang and D.~Wang, ``{Combining Spectral and Spatial Features for Deep
  Learning Based Blind Speaker Separation},'' \emph{IEEE/ACM Trans. Audio,
  Speech, Lang. Process.}, vol.~27, no.~2, pp. 457--468, 2019.

\bibitem{Gannot2017}
S.~Gannot, E.~Vincent \emph{et~al.}, ``{A Consolidated Perspective on
  Multi-Microphone Speech Enhancement and Source Separation},'' \emph{IEEE/ACM
  Trans. Audio, Speech, Lang. Process.}, vol.~25, pp. 692--730, 2017.

\bibitem{Nakatani2010}
T.~Nakatani, T.~Yoshioka \emph{et~al.}, ``{Speech Dereverberation Based on
  Variance-Normalized Delayed Linear Prediction},'' \emph{IEEE Trans. Audio,
  Speech, Lang. Process.}, vol.~18, no.~7, pp. 1717--1731, 2010.

\bibitem{Haeb-Umbach2020}
R.~Haeb-Umbach, J.~Heymann, L.~Drude, S.~Watanabe, M.~Delcroix \emph{et~al.},
  ``{Far-Field Automatic Speech Recognition},'' \emph{Proc. IEEE}, 2020.

\bibitem{Courville2016}
A.~Courville \emph{et~al.}, \emph{Deep Learning}.\hskip 1em plus 0.5em minus
  0.4em\relax MIT Press, 2016.

\bibitem{A.P.Habets2018DereverbBook}
E.~A.~P. Habets and P.~A. Naylor, ``{Dereverberation},'' in \emph{Audio Source
  Separation and Speech Enhancement}, E.~Vincent, T.~Virtanen, and S.~Gannot,
  Eds.\hskip 1em plus 0.5em minus 0.4em\relax Wiley, 2018, pp. 317--343.

\bibitem{Nakatani2019ConvBeamformer}
T.~Nakatani and K.~Kinoshita, ``{A Unified Convolutional Beamformer for
  Simultaneous Denoising and Dereverberation},'' in \emph{IEEE Signal Process.
  Lett.}, vol.~26, no.~6, 2019, pp. 903--907.

\bibitem{Zhao2018cLSTMLateReverb}
Y.~Zhao, D.~Wang, B.~Xu, and T.~Zhang, ``{Late Reverberation Suppression using
  Recurrent Neural Networks with Long Short-Term Memory},'' in \emph{Proc.
  ICASSP}, 2018, pp. 5434--5438.

\bibitem{Zhou2022ShorteningTarget}
R.~Zhou, W.~Zhu, and X.~Li, ``{Single-Channel Speech Dereverberation using
  Subband Network with A Reverberation Time Shortening Target},'' in
  \emph{arXiv preprint arXiv:2204.08765}, 2022.

\bibitem{Quan2022NarrowbandConformer}
C.~Quan and X.~Li, ``{Multichannel Speech Separation with Narrow-band
  Conformer},'' in \emph{Proc. Interspeech}, 2022, pp. 5378--5382.

\bibitem{Hao2021FullSubNet}
X.~Hao, X.~Su, R.~Horaud, and X.~Li, ``{FullSubNet: A Full-Band and Sub-Band
  Fusion Model for Real-Time Single-Channel Speech Enhancement},'' in
  \emph{Proc. ICASSP}, 2021, pp. 6633--6637.

\bibitem{Chen2022FullSubBandplus}
J.~Chen, Z.~Wang, D.~Tuo, Z.~Wu, S.~Kang, and H.~Meng, ``{FullSubNet+: Channel
  Attention FullSubNet with Complex Spectrograms for Speech Enhancement},'' in
  \emph{Proc. ICASSP}, 2022, pp. 7857--7861.

\bibitem{Xu2018GridLSTM}
C.~Xu, W.~Rao, X.~Xiao, E.~S. Chng, and H.~Li, ``{Single Channel Speech
  Separation with Constrained Utterance Level Permutation Invariant Training
  using Grid LSTM},'' in \emph{Proc. ICASSP}, 2018, pp. 6--10.

\bibitem{Liu2020Attn}
Y.~Liu, B.~Thoshkahna, A.~Milani, and T.~Kristjansson, ``{Voice and
  Accompaniment Separation in Music using Self-Attention Convolutional Neural
  Network},'' in \emph{arXiv preprint arXiv:2003.08954}, 2020.

\bibitem{Pandey2021}
A.~Pandey and D.~Wang, ``{Dense CNN with Self-Attention for Time-Domain Speech
  Enhancement},'' \emph{IEEE/ACM Trans. Audio, Speech, Lang. Process.},
  vol.~29, pp. 1270--1279, 2021.

\bibitem{Vaswani2017}
A.~Vaswani \emph{et~al.}, ``{Attention Is All You Need},'' in \emph{Proc.
  NIPS}, 2017.

\bibitem{Wang2021compensation}
Z.-Q. Wang, G.~Wichern, and J.~{Le Roux}, ``{On The Compensation Between
  Magnitude and Phase in Speech Separation},'' \emph{IEEE Signal Process.
  Lett.}, vol.~28, pp. 2018--2022, 2021.

\bibitem{Chen2017a}
Y.~Luo, Z.~Chen, and N.~Mesgarani, ``{Speaker-Independent Speech Separation
  with Deep Attractor Network},'' \emph{IEEE/ACM Trans. Audio, Speech, Lang.
  Process.}, vol.~26, no.~4, pp. 787--796, 2018.

\bibitem{Wisdom2018MixtureConsistency}
S.~Wisdom, J.~R. Hershey, K.~Wilson, J.~Thorpe, M.~Chinen, B.~Patton, and R.~A.
  Saurous, ``{Differentiable Consistency Constraints for Improved Deep Speech
  Enhancement},'' in \emph{Proc. ICASSP}, 2019, pp. 900--904.

\bibitem{Zmolikova2021}
K.~Zmolikova and J.~H. Cernock, ``{BUT System for the First Clarity Enhancement
  Challenge},'' in \emph{Proc. Clarity}, 2021, pp. 1--3.

\bibitem{H.Taal2011}
C.~{H. Taal}, R.~{C. Hendriks} \emph{et~al.}, ``{An Algorithm for
  Intelligibility Prediction of Time-Frequency Weighted Noisy Speech},''
  \emph{IEEE Trans. Audio, Speech, Lang. Process.}, vol.~19, no.~7, pp.
  2125--2136, 2011.

\bibitem{Wang2021seq}
Z.-Q. Wang \emph{et~al.}, ``{Sequential Multi-Frame Neural Beamforming for
  Speech Separation and Enhancement},'' in \emph{Proc. SLT}, 2021, pp.
  905--911.

\bibitem{Wang2022LowLatency}
Z.-Q. Wang, G.~Wichern, S.~Watanabe, and J.~{Le Roux}, ``{STFT-Domain Neural
  Speech Enhancement with Very Low Algorithmic Latency},'' in \emph{IEEE/ACM
  Trans. Audio, Speech, Lang. Process.}, 2022.

\bibitem{Kinoshita2017}
K.~Kinoshita, M.~Delcroix, H.~Kwon, T.~Mori, and T.~Nakatani, ``{Neural
  Network-Based Spectrum Estimation for Online WPE Dereverberation},'' in
  \emph{Proc. Interspeech}, 2017, pp. 384--388.

\bibitem{Cornell2022FB}
S.~Cornell, M.~Pariente \emph{et~al.}, ``{Learning Filterbanks for End-To-End
  Acoustic Beamforming},'' in \emph{Proc. ICASSP}, 2022, pp. 6507--6511.

\bibitem{Kinoshita2016}
K.~Kinoshita, M.~Delcroix \emph{et~al.}, ``{A Summary of The REVERB Challenge:
  State-of-The-Art and Remaining Challenges in Reverberant Speech Processing
  Research},'' \emph{Eurasip J. Adv. Signal Process.}, vol. 2016, no.~1, pp.
  1--19, 2016.

\bibitem{Scheibler2018}
R.~Scheibler, E.~Bezzam, and I.~Dokmanic, ``{Pyroomacoustics: A Python Package
  for Audio Room Simulation and Array Processing Algorithms},'' in \emph{Proc.
  ICASSP}, vol. 2018, 2018, pp. 351--355.

\bibitem{fonseca2020fsd50k}
E.~Fonseca \emph{et~al.}, ``{FSD50K}: An open dataset of human-labeled sound
  events,'' \emph{IEEE/ACM Trans. Audio, Speech, Lang. Process.}, 2021.

\bibitem{Drude2020NARAWPE}
L.~Drude, J.~Heymann, C.~Boeddeker, and R.~Haeb-Umbach, ``{NARA-WPE: A Python
  Package for Weighted Prediction Error Dereverberation in Numpy and Tensorflow
  for Online and Offline Processing},'' in \emph{ITG-Fachtagung
  Sprachkommunikation}, 2020, pp. 216--220.

\bibitem{Yoshioka2015}
T.~Yoshioka, N.~Ito, M.~Delcroix \emph{et~al.}, ``{The NTT CHiME-3 System:
  Advances in Speech Enhancement and Recognition for Mobile Multi-Microphone
  Devices},'' in \emph{Proc. ASRU}, 2015, pp. 436--443.

\bibitem{Heymann2015}
J.~Heymann, L.~Drude, A.~Chinaev, and R.~Haeb-Umbach, ``{BLSTM Supported GEV
  Beamformer Front-End for The 3rd CHiME Challenge},'' in \emph{Proc. ASRU},
  2015, pp. 444--451.

\bibitem{Vincent2006a}
E.~Vincent, R.~Gribonval, and C.~F{\'{e}}votte, ``{Performance Measurement in
  Blind Audio Source Separation},'' \emph{IEEE Trans. Audio, Speech, Lang.
  Process.}, vol.~14, no.~4, pp. 1462--1469, 2006.

\bibitem{Wang2023ICASSPLowLat}
Z.-Q. Wang, S.~Cornell, S.~Choi, Y.~Lee, B.-Y. Kim, and S.~Watanabe, ``{Neural
  Speech Enhancement with Very Low Algorithmic Latency and Complexity via
  Integrated Full- and Sub-Band Modeling},'' in \emph{Proc. ICASSP}, 2023.

\bibitem{ZhangJisi2020}
J.~Zhang, C.~Zorila, R.~Doddipatla, and J.~Barker, ``{On End-to-End
  Multi-Channel Time Domain Speech Separation in Reverberant Environments},''
  in \emph{Proc. ICASSP}, 2020, pp. 6389--6393.

\bibitem{Luo2020e2e}
Y.~Luo, Z.~Chen, N.~Mesgarani, and T.~Yoshioka, ``{End-to-End Microphone
  Permutation and Number Invariant Multi-Channel Speech Separation},'' in
  \emph{Proc. ICASSP}, 2020, pp. 6394--6398.

\bibitem{Taherian2022LBT}
H.~Taherian, K.~Tan, and D.~Wang, ``{Multi-Channel Talker-Independent Speaker
  Separation Through Location-Based Training},'' \emph{IEEE/ACM Trans. Audio,
  Speech, Lang. Process.}, vol.~30, pp. 2791--2800, 2022.

\bibitem{Subakan2022journal}
C.~Subakan, M.~Ravanelli, S.~Cornell \emph{et~al.}, ``{On Using Transformers
  for Speech-Separation},'' \emph{arXiv preprint arXiv:2202.02884}, 2022.

\bibitem{Zhang2021}
J.~Zhang, C.~Zorilă, R.~Doddipatla, and J.~Barker, ``{Time-Domain Speech
  Extraction with Spatial Information and Multi Speaker Conditioning
  Mechanism},'' in \emph{Proc. ICASSP}, 2021, pp. 6084--6088.

\bibitem{Zhang2022BaiduSpeechL3DAS}
G.~Zhang, C.~Wang, L.~Yu, and J.~Wei, ``{Multi-Scale Temporal Frequency
  Convolutional Network with Axial Attention for Multi-Channel Speech
  Enhancement},'' in \emph{Proc. ICASSP}, no.~1, 2022, pp. 9206--9210.

\bibitem{Zhang2022Axial}
G.~Zhang, L.~Yu, C.~Wang, and J.~Wei, ``{Multi-Scale Temporal Frequency
  Convolutional Network With Axial Attention for Speech Enhancement},'' in
  \emph{Proc. ICASSP}, 2022, pp. 9122--9126.

\bibitem{Li2022PCGL3DAS}
J.~Li, Y.~Zhu, D.~Luo, Y.~Liu, G.~Cui, and Z.~Li, ``{The PCG-AIID System for
  L3DAS22 Challenge: MIMO and MISO Convolutional Recurrent Network for Multi
  Channel Speech Enhancement and Speech Recognition},'' in \emph{Proc. ICASSP},
  2022, pp. 9211--9215.

\bibitem{ren2021neural}
X.~Ren, L.~Chen \emph{et~al.}, ``A neural beamforming network for {B-Format}
  {3D} speech enhancement and recognition,'' in \emph{Proc. MLSP}, 2021.

\bibitem{Cornell2022Clarity}
S.~Cornell, Z.-Q. Wang, Y.~Masuyama, S.~Watanabe, M.~Pariente, and N.~Ono,
  ``{Multi-Channel Target Speaker Extraction with Refinement: The WAVLAB
  Submission to The Second Clarity Enhancement Challenge},'' in \emph{Proc.
  Clarity}, 2022, pp. 1--3.

\bibitem{cornell2023multi}
S.~Cornell, Z.-Q. Wang, Y.~Masuyama, S.~Watanabe, M.~Pariente, N.~Ono, and
  S.~Squartini, ``{Multi-Channel Speaker Extraction with Adversarial Training:
  The WAVLAB Submission to The Clarity ICASSP 2023 Grand Challenge},'' in
  \emph{Proc. ICASSP}, 2023, pp. 1--2.

\end{thebibliography}

\noindent Authors’ photographs and biographies not available at the time of publication.

\end{document}